\documentclass[aip,pop,reprint,twocolumn,showpacs,floatfix,preprintnumbers,amsmath,amssymb,nofootinbib]{revtex4-1}

\usepackage{graphicx}
\usepackage{dcolumn}
\usepackage{bm}
\usepackage{wasysym}
\usepackage{xfrac}
\usepackage{nicefrac}
\usepackage{courier}
\usepackage{dcolumn}
\usepackage{bigints}
\usepackage{gensymb}
\usepackage{hyperref}
\usepackage{fancyhdr}

\makeatletter
\def\frontmatter@makefnmark{%
 \@textsuperscript{%
  \normalfont\@thefnmark
 }%
}%
\makeatother

\begin{document}


\title{A semi-analytic model of magnetized liner inertial fusion$^\dagger$}
\fancyhf{}
\rhead{Page \thepage\ of 24}
\cfoot{$^\dagger$Journal Reference: Phys. Plasmas {\bf 22}, 052708 (2015); \url{http://dx.doi.org/10.1063/1.4918953}.}
\renewcommand{\headrulewidth}{0.0pt}

\author{Ryan~D.~McBride}
\noaffiliation
\affiliation{Sandia National Laboratories, Albuquerque, New Mexico 87185, USA}
\author{Stephen~A.~Slutz}
\noaffiliation
\affiliation{Sandia National Laboratories, Albuquerque, New Mexico 87185, USA}

\date{\today}

\begin{abstract}
Presented is a semi-analytic model of magnetized liner inertial fusion (MagLIF). This model accounts for several key aspects of MagLIF, including: (1) preheat of the fuel (optionally via laser absorption); (2) pulsed-power-driven liner implosion; (3) liner compressibility with an analytic equation of state, artificial viscosity, internal magnetic pressure, and ohmic heating; (4) adiabatic compression and heating of the fuel; (5) radiative losses and fuel opacity; (6) magnetic flux compression with Nernst thermoelectric losses; (7) magnetized electron and ion thermal conduction losses; (8) end losses; (9) enhanced losses due to prescribed dopant concentrations and contaminant mix; (10) deuterium-deuterium and deuterium-tritium primary fusion reactions for arbitrary deuterium to tritium fuel ratios; and (11) magnetized $\alpha$-particle fuel heating. We show that this simplified model, with its transparent and accessible physics, can be used to reproduce the general 1D behavior presented throughout the original MagLIF paper [S.~A.~Slutz {\it et al}., Phys. Plasmas {\bf 17}, 056303 (2010)].  We also discuss some important physics insights gained as a result of developing this model, such as the dependence of radiative loss rates on the radial fraction of the fuel that is preheated.
\end{abstract}

\pacs{52.58.Lq, 84.70.+p}

\keywords{magnetized liner inertial fusion, MagLIF, Z machine, Z accelerator, Z300, Z800, Z beamlet laser, ZBL, pulsed power, fusion, z-pinch, inertial confinement fusion, ICF, magneto-inertial fusion, MIF}

\maketitle
\thispagestyle{fancy}

\section{\label{sec:intro}Introduction}

The Magnetized Liner Inertial Fusion (MagLIF) concept\cite{Slutz_PoP_2010,Slutz_PRL_2012} is presently being investigated experimentally\cite{Sinars_PRL_2010,Sinars_MRT_PoP_2010,Cuneo_IEEE-TPS_2012,Lemke_SCCM_2011,Martin_SCCM_2011,Martin_PoP_2012,McBride_PRL_2012,McBride_PoP_2013,Dolan_RSI_2013,Peterson_PoP_2012,Peterson_PoP_2013,Peterson_PRL_2014,Awe_PRL_2013,Awe_PoP_2014,Sefkow_PoP_2014,Gomez_PRL_2014,Schmit_PRL_2014,Knapp_PoP_2015} using the Z facility\cite{Rose_PRSTAB_2010_EM_model,Savage_PPC_2011} at Sandia National Laboratories. MagLIF is part of a broader class of concepts referred to collectively as magneto-inertial fusion (MIF).\cite{Khariton_UFN_1976,Khariton_SPU_1976,Mokhov_SPD_1979,Sweeney_NF_1981,Lindemuth_PoF_1981,Lindemuth_NF_1983,Jones_NF_1986,Hasegawa_PRL_1986,Lindemuth_PRL_1995,Kirkpatrick_FT_1995,Degnan_PRL_1999,Siemon_CPPCF_1999,Basko_NF_2000,Kemp_NF_2001,Ryutov_CPPCF_2001,Kemp_NF_2003,Intrator_IEEE-TPS_2004,Garanin_IEEE-MegaGaussConf_2006}  These concepts seek to significantly reduce the implosion velocity and pressure requirements of traditional inertial confinement fusion (ICF)\cite{Nuckolls_Nature_1972,Lindl_PoP_1995,Perkins_PRL_2009,Meezan_PoP_2010,Hurricane_Nature_2014} by using a magnetic field to thermally insulate the hot fuel\cite{Landshoff_PR_1949} from a cold pusher and to increase fusion product confinement.

The MagLIF concept at Sandia uses the electromagnetic pulse supplied by the Z accelerator to radially implode an initially solid cylindrical metal tube (liner) filled with preheated and premagnetized fusion fuel (deuterium or deuterium-tritium). The implosion is a result of the fast z-pinch process, where a large gradient in the applied magnetic field pressure operates near the liner's outer surface.\cite{Ryutov_FZP_RMP_2000,Cuneo_IEEE-TPS_2012}  One- and two-dimensional simulations of MagLIF using the LASNEX radiation magnetohydrodynamics code \cite{Zimmerman_CPPCF_1975} predict that if sufficient liner integrity can be maintained throughout the implosion, then significant fusion yield ($>$100 kJ) can be attained on the Z accelerator when deuterium-tritium fuel is used and the accelerator's Marx generators are charged to 95 kV to obtain a peak drive current of about 27 MA.\cite{Slutz_PoP_2010,Cuneo_IEEE-TPS_2012}

To maintain liner integrity throughout the implosion, liners with thick walls have been proposed\cite{Slutz_PoP_2010,Slutz_PRL_2012} and used experimentally.\cite{Sinars_PRL_2010,Sinars_MRT_PoP_2010,McBride_PRL_2012,McBride_PoP_2013,Awe_PRL_2013,Awe_PoP_2014,Gomez_PRL_2014} Thick walls mitigate the deleterious effects of the Magneto-Rayleigh-Taylor (MRT) instability,\cite{Harris_MRT_PoF_1962,Ott_MRT_PRL_1972,Ryutov_FZP_RMP_2000,Reinovsky_IEEE_2002,Miles_MRT_PoP_2009,Sinars_PRL_2010,Sinars_MRT_PoP_2010,Lau_PRE_2011,Zhang_PoP_2012,McBride_PRL_2012,Cuneo_IEEE-TPS_2012,McBride_PoP_2013,Awe_PRL_2013,Awe_PoP_2014} which first develops on the liner's outer surface, and then works its way inward, toward the liner's inner/fuel-confining surface, throughout the implosion. The parameter that is typically used to describe the robustness of a liner to the MRT instability is the liner's initial aspect ratio
\begin{equation}\label{Ar}
A_{r0}\equiv \frac{r_{l0}}{\delta_{l0}},
\end{equation}
where $r_{l0}$ is the liner's initial outer radius and $\delta_{l0}$ is the liner's initial wall thickness. Lower $A_{r0}$ liners are more robust to the MRT instability, but their use results in slower implosion velocities; thus, there is a tradeoff between liner robustness and implosion efficiency.

Two-dimensional LASNEX simulations predict that liners with $A_{r0}<10$ should be robust enough to keep the MRT instability from overly disrupting the fusion burn at stagnation.\cite{Slutz_PoP_2010} These preliminary LASNEX simulations predict a broad optimum in the fusion yield surrounding $A_{r0}\approx 6$ (see Fig.~10 in Ref.~\onlinecite{Slutz_PoP_2010}). Experiments\cite{McBride_PRL_2012,McBride_PoP_2013} on the Z accelerator with no initial axial magnetic field have found the implosion dynamics of $A_{r0}=6$ Be liners to be in good agreement with the 2D LASNEX predictions found in the original MagLIF paper\cite{Slutz_PoP_2010} (e.g., compare Fig.~2 in Ref.~\onlinecite{McBride_PoP_2013} with Fig.~9 in Ref.~\onlinecite{Slutz_PoP_2010}). By contrast, $A_{r0}=6$ Be liner experiments on the Z accelerator that included an initial axial magnetic field of about 10 T have revealed the presence of a helical instability structure that is fundamentally 3D in nature,\cite{Awe_PRL_2013,Awe_PoP_2014} and thus could not be captured by the 2D LASNEX simulations; regardless, the inclusion of the initial axial field seems only to have improved overall implosion stability.\cite{Awe_PoP_2014}  Moreover, the first fully-integrated experimental tests of MagLIF have shown that a fuel implosion driven by an $A_{r0}=6$ Be liner, combined with preheat and premagnetization, does indeed result in fusion relevant conditions on the Z accelerator.\cite{Gomez_PRL_2014,Schmit_PRL_2014,Sefkow_PoP_2014}  Therefore, we assume that the qualitative 1D versus 2D behavior shown in Fig.~10 of Ref.~\onlinecite{Slutz_PoP_2010} holds fairly true for $A_{r0}\lesssim 6$, and thus that MagLIF can be modeled reasonably well in 1D for $A_{r0}\lesssim 6$.

The leading hypothesis for what seeds the MRT development is the electrothermal instability (ETI).\cite{Peterson_PoP_2012,Peterson_PoP_2013,Peterson_PRL_2014} Recent experiments have shown that ETI growth can be tamped by applying a thick dielectric coating to the outer surface of the metal liner, thereby delaying the onset of nonlinear MRT growth.\cite{Peterson_PRL_2014}  Thus, it is hopeful that the application of dielectric coatings may enable MagLIF's 2D and/or 3D $\approx$ 1D performance to be extended to liners with aspect ratios significantly greater than 6.  Note that these assumptions do not account for possible azimuthal variations; the effects of azimuthal variations on fuel confinement are beyond the scope of this paper, but they are presently being investigated in other studies.\cite{Jennings_personal_2015,Sefkow_personal_2015} Presumably, the deleterious effects of azimuthal variations are also mitigated by the use of low $A_{r0}$ liners, and perhaps by dielectric coatings as well.

Since MagLIF calls for low $A_{r0}$ liners to mitigate the MRT instability, the resulting liner implosions are too slow to shock heat the fuel (i.e., the fuel heating due to compression is approximately adiabatic). Thus, to heat the fuel to fusion-relevant temperatures ($>$1 keV) while simultaneously lowering fuel convergence requirements, the MagLIF concept calls for preheating the fuel just prior to the implosion. The parameter typically used to describe fuel convergence is the convergence ratio
\begin{equation}\label{Crp}
C_{r}(t)\equiv \frac{r_{g0}}{r_{g}(t)},
\end{equation}
where $r_{g0}$ and $r_{g}(t)$ are the initial and time-dependent radii of the fuel-liner interface, respectively. The convergence ratio is often evaluated at particular times of interest, such as the time of peak fusion power, $C_{rp}$, and the time of minimum $r_g$, i.e., the time of ``bounce", $C_{rb}$.  In many cases, $C_{rp}$ and $C_{rb}$ have fairly similar values, but differences between these two parameters can indicate important phenomena such as fuel collapse or fuel ignition.  To reduce $C_{rp}$ and $C_{rb}$ requirements to $\lesssim 30$, the MagLIF concept calls for preheat temperatures of $\gtrsim 100$ eV.

At the Z facility, fuel preheating has been accomplished using the Z beamlet laser (ZBL),\cite{Rambo_AO_2005,Gomez_PRL_2014} which was originally coupled to the Z pulsed-power accelerator for diagnostic purposes (e.g., radiography\cite{Bennett_2frame_RSI_2008,Sinars_RSI_2004}). In the first fully-integrated MagLIF experiments,\cite{Gomez_PRL_2014} ZBL provided about 2 kJ of 532-nm light in a 2-ns pulse.  Recently, ZBL has been upgraded to deliver about 4 kJ in a 4-ns pulse, and plans are in place to further increase ZBL's delivered pulse energy to 6 kJ and beyond over the next several years.

To keep the fuel hot during the relatively slow implosion of a MagLIF liner, the concept requires premagnetization of the fuel using an axially-aligned $B_{z}$ field. About 10--50 T are to be supplied by external $B_{z}$ coils. This initial seed field is to be amplified by a factor of 10$^2$--10$^3$ within the fuel-filled volume of the imploding liner via magnetic flux compression. This large axial field is required to mitigate energy loss from the fuel due to electron and ion thermal conduction. Additionally, the axial field should enhance $\alpha$-particle confinement and heating of the fuel, and thus increase the overall fusion yield.

In the first fully-integrated MagLIF experiments,\cite{Gomez_PRL_2014} as well as in liner implosion dynamics experiments,\cite{Awe_PRL_2013,Awe_PoP_2014} initial seed fields of up to 10 T have been supplied by a newly developed applied $B$ on Z (ABZ) subsystem at the Z facility.\cite{Rovang_RSI_2014}  Analysis of the neutron diagnostics data collected during the first fully-integrated MagLIF experiments indicates that significant flux compression did indeed occur.\cite{Schmit_PRL_2014,Knapp_PoP_2015}  Also, the ABZ capabilities at the Z facility have recently been upgraded to 15 T, and plans are in place to further increase these fields to 30 T in the next couple of years.\cite{Rovang_RSI_2014}

To elucidate some of the key physics issues relevant to MagLIF, we have developed a new semi-analytic model of the concept. This model is formulated as a system of ordinary differential equations (ODEs) that are straightforward to solve, particularly with standard software tools such as MATLAB\textsuperscript{\textregistered}, IDL\textsuperscript{\textregistered}, Mathematica\textsuperscript{\textregistered}, etc.  This model accounts for: (1) preheat of the fuel (optionally via laser absorption); (2) pulsed-power-driven liner implosion; (3) liner compressibility with an analytic equation of state, artificial viscosity, internal magnetic pressure, and ohmic heating; (4) adiabatic compression and heating of the fuel; (5) radiative losses and fuel opacity; (6) magnetic flux compression with Nernst thermoelectric losses; (7) magnetized electron and ion thermal conduction losses; (8) end losses; (9) enhanced losses due to prescribed dopant concentrations and contaminant mix; (10) deuterium-deuterium and deuterium-tritium primary fusion reactions for arbitrary deuterium to tritium fuel ratios; and (11) magnetized $\alpha$-particle heating. This model has been implemented in a code called SAMM (Semi-Analytic MagLIF Model).  Simulations using SAMM typically take about 30 seconds to run on a laptop using the ode23 solver in MATLAB\textsuperscript{\textregistered}.  Using the parallel computing cluster at Sandia, parameter scans of about 2000 simulations can be completed in as little as 10 minutes.

Semi-analytic models have proven themselves useful in the past.\cite{Lindemuth_NF_1983,Dahlin_PS_2004} Their fast run times allow the system parameter space to be explored rapidly.  Also, the physics included in semi-analytic models are often transparent and thus accessible by anyone, which is not always the case for more sophisticated codes (e.g., the LASNEX code\cite{Zimmerman_CPPCF_1975} that the original MagLIF paper\cite{Slutz_PoP_2010} was based on).  Finally, one of the most important aspects of developing simplified models such as SAMM, is that it forces those involved to verify and understand the results generated by more sophisticated codes.  For example, when differences are observed between a simple model and a more advanced code, can those differences be explained in terms of the simplifying assumptions made in the reduced model?  This type of questioning often leads to new physics insights that would otherwise be overlooked. For example, during the development of SAMM, we found an important relationship between radiative loss rates and the initial radial fraction of the fuel that is preheated; this finding and others are discussed further in this paper.

The remainder of this paper is organized as follows. In Sec. \ref{sec:model}, we present our semi-analytic model of MagLIF. In Sec.~\ref{sec:results}, we verify this model relative to the 1D results presented in the original MagLIF paper (i.e., Ref.~\onlinecite{Slutz_PoP_2010}). In Sec. \ref{sec:summary}, we summarize this work. Unless otherwise specified, all units are SI. 

\section{\label{sec:model}A semi-analytic model of MagLIF}
\subsection{\label{sec:circuit}Overview of the model and the radial distribution of the driving azimuthal magnetic field}

\begin{figure}
\includegraphics[width=\columnwidth]{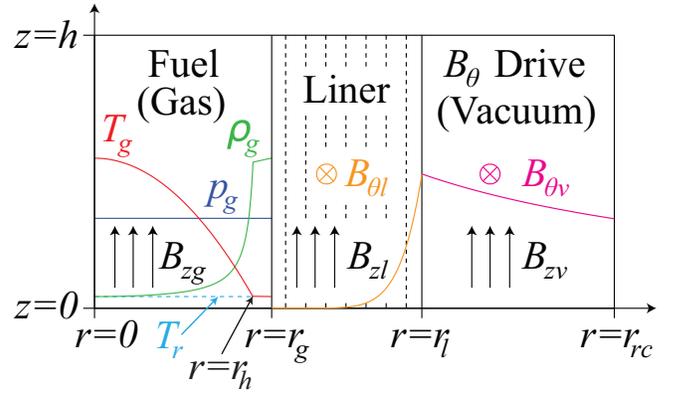}
\caption{\label{fig:schematic} Schematic overview of the semi-analytic MagLIF model. There are three primary regions: a fuel region, a liner region, and a vacuum region.  The system height is $h$; the thermally-insulating axial magnetic field, which is initially distributed uniformly over all regions, is $B_{z}$; the radius of the fuel-liner interface is $r_g$; the liner's outer radius is $r_l$; the return current radius is $r_{rc}$; the azimuthal magnetic field, which drives the cylindrical implosion, is $B_{\theta}$.  Normalized profiles are shown for $B_{\theta}$ in the vacuum region, $B_{\theta v}$ (magenta), and in the liner region,  $B_{\theta l}$ (orange); their analytic expressions are given by Eqs.~\ref{Bthetav} and \ref{Bthetal} in the text ($B_{\theta}$ is assumed to be zero in the fuel region).  The liner region is further divided into multiple concentric liner shells; this discretization is necessary to avoid overdriving the fuel.  Within the fuel region, normalized profiles are shown for the gas pressure, $p_g$ (blue), the gas temperature, $T_g$ (red), the gas density, $\rho_g$ (green), and the radiation temperature, $T_r$ (cyan). The pressure profile is flat throughout the fuel (i.e., we have made an isobaric assumption due to the subsonic nature of MagLIF implosions). The gas temperature and density profiles thus have an inverse dependence to one another; their analytic expressions are given by Eqs.~\ref{Tgr}--\ref{rhogrshelf} in the text.  The radiation temperature is nearly constant across the fuel region.  The fuel region is further divided into a hot spot region from $r=0$ to $r_h$ and a cold dense shelf region from $r_h$ to $r_g$. The gas temperature in the shelf region is equal to the radiation temperature (i.e., the fuel material and the radiation field are in thermodynamic equilibrium in the shelf region).  The shelf region erodes away throughout the implosion, until $r_h=r_g$, due to thermal transport from the hot spot to the shelf.  The shelf region is only present if the fuel is preheated from $r=0$ to $r<r_g$.}
\end{figure}

An overview of the semi-analytic MagLIF model is provided in Fig.~\ref{fig:schematic}.
The liner implosion, and thus the fuel implosion, is driven by the pressure associated with the azimuthal magnetic field, which is supplied by the pulsed-power driver (e.g., the Z accelerator).  Because of the cylindrical symmetry, the azimuthal field in the vacuum region is
\begin{equation}\label{Bthetav}
B_{\theta v}(r)=\frac{\mu_0I_l}{2\pi r},
\end{equation}
where $\mu_0=4\pi\times 10^{-7}$ H/m is the permeability of free space and $I_l$ is the liner current. We assume that $B_\theta$ is partially diffused into the liner wall, and that its distribution in the liner region can be described by
\begin{equation}\label{Bthetal}
B_{\theta l}(r)=\frac{\mu_0I_l}{2\pi r_l}\left( \frac{r-r_g}{r_l-r_g} \right)^\beta,
\end{equation}
where the constant power $\beta$ is found by forcing the amplitude of $B_{\theta l}(r)$ to drop by one $e$-folding within one skin depth, $\delta_{skin}$, of the liner's outer surface; that is, we set
\begin{equation}
B_{\theta l}(r_l-\delta_{skin})= \frac{\mu_0I_l}{2\pi r_l} \cdot \frac{1}{e},
\end{equation}
where $e=2.71828$ is the base of the natural logarithm, and solve for $\beta$, which gives
\begin{equation}\label{beta}
\beta=\max\left\{1,\left|\frac{\ln(1/e)}{\ln\left(\frac{\delta_{l0}-\delta_{skin}}{\delta_{l0}}\right)}\right|\right\}.
\end{equation}
The skin depth is given by
\begin{equation}\label{skin}
\delta_{skin}=\sqrt{\frac{4 \rho_e \tau_r}{\pi\mu_0}},
\end{equation}
where $\rho_e$ is the initial electrical resistivity of the liner material, and $\tau_r$ is the rise time of the driving $B_\theta$ pulse (e.g., $\sim 130$ ns for the Z accelerator).  Some parameters of interest are provided in Table~\ref{rhoe}. The purpose of assuming a simple power-law dependence for $B_{\theta l}(r)$, rather than an exponential dependence, is that it allows us to integrate both the total magnetic flux and the total magnetic field energy analytically, which enables us to obtain simple analytic expressions for the inductance, circuit response, and ohmic dissipation due to $B_{\theta l}$.
\begin{table}
\caption{\label{rhoe}Parameters for $B_{\theta l}(r)$ (where $\tau_r=130$ ns).}
\begin{ruledtabular}
\begin{tabular}{cccc}
 Material & $\rho_{e}$ [n$\Omega$$\cdot$m] & $\delta_{skin}$ [$\mu$m] & $\beta$ \\
\hline
Li & 92.8 & 110.6 & 3.683 \\
Be & 36 & 68.9 & 6.239 \\
Al & 28.2 & 60.9 & 7.118 \\
\end{tabular}
\end{ruledtabular}
\end{table}

\subsection{\label{sec:circuit}Circuit model for the Z pulsed-power driver}

\begin{figure}
\includegraphics[width=\columnwidth]{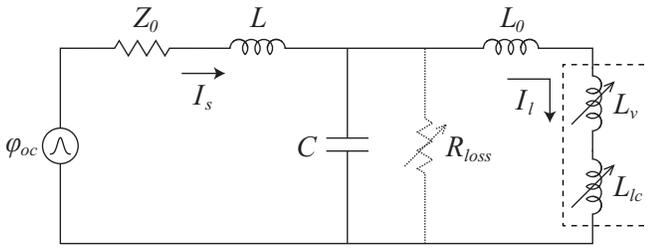}
\caption{\label{fig:LRCdrive} An equivalent circuit model for the Z pulsed-power accelerator, where $\varphi_{oc}(t)$ is the open-circuit voltage used to drive the simulation (see Fig.~\ref{fig:Voc}), $Z_0$=0.18 $\Omega$, $L$=8.34 nH, $C$=8.41 nF, and $L_0$$\approx$5 nH.  The elements $L_v(t)$ and $L_{lc}(t)$ are given by Eqs.~\ref{Lv} and \ref{Llc} in the text and they represent the coupling of the circuit to the dynamic volume of the model illustrated in Fig.~\ref{fig:schematic}.}\end{figure}

\begin{figure}
\includegraphics[width=\columnwidth]{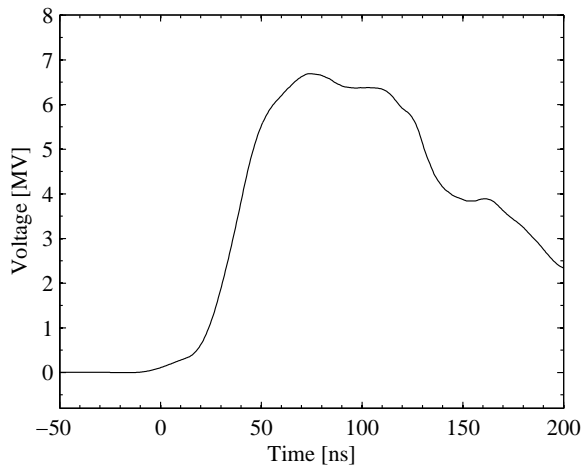}
\caption{\label{fig:Voc} Example open-circuit voltage waveform for the Z pulsed-power accelerator. This waveform was constructed using twice the forward-going voltage measured at the Z accelerator's vacuum-insulator stack.}
\end{figure}

The liner current in Eqs.~\ref{Bthetav} and \ref{Bthetal} can be specified to drive the simulation directly, or it can be derived from a circuit model. For the circuit model option, we use the circuit illustrated in Fig.~\ref{fig:LRCdrive}, which has been shown to adequately represent the Z accelerator coupled to z-pinch loads.\cite{McBride_PRSTAB_2010} With this model, simulations are driven by an open-circuit voltage, $\varphi_{oc}(t)$. This voltage is twice the forward-going voltage at the vacuum-insulator stack on Z, and it can be constructed from the data of previous Z experiments using Eq.~4 in Ref.~\onlinecite{McBride_PRSTAB_2010}. An example $\varphi_{oc}(t)$ waveform for Z is shown in Fig. \ref{fig:Voc}.

This circuit model also requires $R_{loss}(t)$ to be specified. This is a resistance used to model shunt current losses on Z, and it can be derived from previous Z experiment data using the methods discussed in Refs.~\onlinecite{Jennings_IEEE_2010,McBride_PRSTAB_2010,Gomez_PPC_2013}. For simplicity, we will take $R_{loss}(t)$ to be large and constant ($>$10 $\Omega$) throughout this paper, which essentially eliminates current loss from the results presented herein.

To solve the circuit shown in Fig.~\ref{fig:LRCdrive}, we begin by writing the voltage across the capacitor
\begin{equation}
\label{Voc}
\varphi_{c} = \varphi_{oc} - Z_0I_{s} - L\dot{I}_{s},
\end{equation}
and thus
\begin{equation}
\label{Isdot}
\dot{I}_s = \frac{\varphi_{oc} - Z_0I_{s} - \varphi_c}{L}.
\end{equation}
The liner current is
\begin{equation}
\label{Il}
I_l = I_s-C\dot{\varphi}_c-\varphi_c/R_{loss},
\end{equation}
and thus
\begin{equation}
\label{Vcdot}
\dot{\varphi}_c = \frac{I_s-I_l-\varphi_c/R_{loss}}{C}.
\end{equation}
The voltage required to supply $B_\theta$ flux to the model's volume illustrated in Fig.~\ref{fig:schematic} is
\begin{equation}
\label{Vrc1}
\varphi_{rc} = \varphi_c - L_0\dot{I}_l.
\end{equation}
Since the axial current density in the liner wall goes to zero at $r=r_g$, which is implied by our assumption for $B_{\theta l}(r)$, then by the integral form of Faraday's law
\begin{equation}
\label{faraday}
\oint_c {\bf E \;\cdot\;} d{\bf l} = -\dot{\Phi}_\theta,
\end{equation}
where the path integral curve $c$ is around the model's volume illustrated in Fig.~\ref{fig:schematic}, ${\rm\bf E}$ is the electric field vector, $d{\bf l}$ is an infinitesimal path element vector along curve $c$, and ${\Phi}_{\theta}$ is the total $B_\theta$ flux in the model's volume illustrated in Fig.~\ref{fig:schematic}, we know that we can write $\varphi_{rc}$ also as
\begin{equation}
\label{Vrc2}
\varphi_{rc} = \dot{\Phi}_{\theta} = \dot{\Phi}_{\theta v}+\dot{\Phi}_{\theta l},
\end{equation}
where ${\Phi}_{\theta v}$ and ${\Phi}_{\theta l}$ are the total $B_\theta$ fluxes in the vacuum and liner regions, respectively. For the vacuum region, we have
\begin{equation}
\label{Phi_theta_v_dot}
\dot{\Phi}_{\theta v} = L_v\dot{I}_l+\dot{L}_v I_l,
\end{equation}
where $L_v$ is the standard coaxial vacuum inductance\cite{Cheng}
\begin{equation}\label{Lv}
L_v(t)=\frac{\mu_0h}{2\pi}\ln\left[\frac{r_{rc}}{r_l(t)}\right],
\end{equation}
and thus
\begin{equation}\label{Lvdot}
\dot{L}_v(t)=-\frac{\mu_0h}{2\pi}\left(\frac{\dot{r}_l}{r_l}\right).
\end{equation}
For the liner region, we have
\begin{equation}
\Phi_{\theta l}=h\int_{r_g}^{r_l}B_{\theta l}(r)\cdot dr = \frac{\mu_0 h I_l (r_l-r_g)}{2\pi r_l(\beta+1)},
\end{equation}
and thus
\begin{equation}\label{Phi_theta_liner_dot}
\dot{\Phi}_{\theta l}= \frac{\mu_0 h}{2\pi (\beta+1)}\left[\dot{I}_l\left(1-\frac{r_g}{r_l}\right)+I_l\left(\frac{r_g\dot{r}_l}{r_l^2}-\frac{\dot{r}_g}{r_l}\right)\right].
\end{equation}
It is convenient to define an effective inductance that can be used to describe the circuit response due to the $B_\theta$ flux in the liner region.  From Eq.~\ref{Phi_theta_liner_dot}, we define this effective inductance as
\begin{equation}\label{Llc}
L_{lc}(t)\equiv \frac{\mu_0h}{2\pi(\beta +1)}\left(1-\frac{r_g}{r_l}\right),
\end{equation}
and thus
\begin{equation}\label{Llcdot}
\dot{L}_{lc}(t)= \frac{\mu_0h}{2\pi(\beta +1)}\left(\frac{r_g \dot{r}_l}{r_l^2}-\frac{\dot{r}_g}{r_l}\right).
\end{equation}
With these expressions, Eq.~\ref{Phi_theta_liner_dot} can be rewritten as
\begin{equation}
\label{Phi_theta_liner_dot_2}
\dot{\Phi}_{\theta v} = L_{lc}\dot{I}_l+\dot{L}_{lc} I_l.
\end{equation}
Finally, combining Eqs.~\ref{Vrc1}, \ref{Vrc2}, \ref{Phi_theta_v_dot}, and \ref{Phi_theta_liner_dot_2} and solving for $\dot{I}_l$ gives
\begin{equation}
\label{Ildot}
\dot{I}_l=\frac{\varphi_c-I_l(\dot{L}_v + \dot{L}_{lc})}{L_0+L_v+L_{lc}}.
\end{equation}

\subsection{\label{sec:circuit}Drive energetics and ohmic liner heating}

It is important to realize that the definition for $L_{lc}$ above is not the same as the definition for standard inductance; the standard inductance is found from the total magnetic field energy in the liner region\cite{Cheng}
\begin{equation}\label{EBthetal}
E_{B_{\theta l}} = \frac{1}{2}L_{l}I_l^2=h\int_{r_g}^{r_l}\frac{B_{\theta l}^2}{2\mu_0}\cdot 2\pi r \cdot dr,
\end{equation}
which gives
\begin{equation}\label{Ll}
L_l(t)=\frac{\mu_0 h (2\beta r_l+r_l+r_g)(r_l-r_g)}{4\pi r_l^2(\beta +1)(2\beta +1)},
\end{equation}
and thus
\begin{equation}\label{Ll}
\dot{L}_l(t)=-\frac{\mu_0 h (r_g+\beta r_l)(\dot{r}_g r_l - r_g\dot{r}_l)}{2\pi r_l^3(\beta +1)(2\beta +1)}.
\end{equation}
Note that with this standard inductance, $\dot{\Phi}_{\theta l}\neq L_l\dot{I_l}+\dot{L_l}I_l$.  Differences exist between $L_{lc}$ and $L_l$ because the liner current is distributed radially across the liner region.  The distributed current is a result of our assumed $B_{\theta l}(r)$, which we chose in an attempt to model the magnetic diffusion and ohmic dissipation known to occur in thick-walled liner implosions.\cite{diffusion}

These two definitions for inductance provide us with a clean and simple way to express the following powers being delivered to the volume of Fig.~\ref{fig:schematic}: (a) the total electromagnetic power, $P_{EM}$; (b) the power going into just the azimuthal magnetic field, $P_{B_{\theta}}$; (c) the power going into the kinetic motion and compression of the fuel, liner, and axial magnetic field, $P_{kin}$; and (d) the ohmic heating power, $P_{\Omega}$.

The total electromagnetic power is the sum $P_{EM} =P_{B_{\theta}}+P_{kin}+P_{\Omega}$; it can also be expressed in terms of the drive circuit as
\begin{equation}\label{PEM}
P_{EM} =\dot{\Phi}_{\theta} I_l = (L_v+L_{lc})\dot{I}_l I_l + (\dot{L}_v + \dot{L}_{lc})I_l^2.
\end{equation}
From Eq.~\ref{EBthetal}, the power going into just the $B_\theta$ field is
\begin{equation}\label{PBthetal}
P_{B_{\theta}} = \dot{E}_{B_{\theta}} = (L_v+L_l) \dot{I}_l I_l + \frac{1}{2}(\dot{L}_v+\dot{L}_l) I_l^2.
\end{equation}
To obtain an expression for $P_{kin}$, we consider the case of a perfectly conducting liner, where $P_{\Omega}=0$, and $L_{lc} = L_{l}=0$ because the liner current is a surface current at $r=r_l$ (i.e., $\beta \to \infty$).\cite{f0} This gives
\begin{equation}\label{PkinPC}
P_{kin}(\beta \to \infty) = P_{EM} - P_{B_{\theta}} = \frac{1}{2}\dot{L}_v I_l^2,
\end{equation}
which is an upper bound for $P_{kin}$.  From numerical tests with finite conductivity, we find that to within a few percent over most of the implosion
\begin{equation}\label{Pkinapprox}
P_{kin}\approx P_{kin}(\beta \to \infty) = \frac{1}{2}\dot{L}_v I_l^2.
\end{equation}
(Below, in Eq.~\ref{Pkinexact}, we provide a more precise definition and explanation of $P_{kin}$.) Finally, the power associated with ohmic dissipation is found from
\begin{equation}\label{Pohmexact}
P_{\Omega}=\left| P_{EM}-P_{B_{\theta}}-P_{kin}\right|.
\end{equation}
Substituting Eqs.~\ref{PEM}, \ref{PBthetal}, and \ref{Pkinapprox} into Eq.~\ref{Pohmexact} gives the approximate (and lower bound) solution
\begin{equation}\label{Pohmapprox}
P_{\Omega} \approx \left| (L_{lc}-L_{l}) \dot{I}_l I_l + \left(\dot{L}_{lc}-\frac{1}{2}\dot{L}_l\right)I_l^2 \right|.
\end{equation}

Note that Eqs.~\ref{PEM}, \ref{Pohmexact}, and \ref{Pohmapprox} are valid only if the pulsed-power generator can supply azimuthal flux to the liner wall faster than the flux can be dissipated ohmically, and thus $\delta_{skin}< r_l-r_g$.  In typical MagLIF experiments, this condition is essentially always met due to the use of electrically-conductive, thick-walled liners, and because the pulsed liner current changes rapidly with time throughout almost the entire experiment.\cite{uniformJz,Knoepfel}  In numerical tests, we have found that Eqs.~\ref{Pohmexact} and \ref{Pohmapprox} generate ohmic dissipation rates that agree well with those calculated by full radiation magnetohydrodynamics simulations.

Accounting for $P_{\Omega}$ is important because without it, the liner region remains cooler and more compressible than expected throughout the implosion (i.e., on a ``lower adiabat").  Upon stagnation, an overly compressed liner region will drive the fuel to a pressure that is too high, and thus result in an overly optimistic fusion yield.

\subsection{\label{sec:circuit}Liner dynamics and compression}

To account for liner compressibility, we first divide the liner region into $N_{ls}\gtrsim 20$ concentric thin liner shells and write an equation of motion for the interface between each shell (internal interfaces), as well as for the fuel-liner interface and for the liner-vacuum interface (external interfaces). To each internal interface, we assign a mass of $m_{ls}=m_l/N_{ls}$, where $m_l$ is the total liner mass; to each external interface, we assign a mass of $m_{ls}/2$.  There are $N_{li}=N_{ls}+1$ liner interfaces in total.  The radial positions of the liner interfaces, $r_{l,i}$, are distributed from $r_g\equiv r_{l,i=1}$ to $r_l\equiv r_{l,i=N_{li}}$.  The initial radial position of interface $i+1$ is given by
\begin{multline}
r_{l,i+1}(t_0)=\sqrt{r_{l,i}^2(t_0)+m_{ls}/(\pi h \rho_{l0})} \\ \left(i=1,2,\ldots,N_{li}-2\right),
\end{multline}
where $\rho_{l0}$ is the initial mass density of the liner material.  The center of mass radial position for liner shell $s$ is given by
\begin{multline}\label{rls}
r_{l,s}(t)=\sqrt{\left[r_{l,i=s}^2(t)+r_{l,i=s+1}^2(t)\right]/2} \\ \left(s=1,2,\ldots,N_{ls}\right).
\end{multline}
Since there are $N_{li}$ interfaces, there are $N_{li}$ equations of motion.  The equations of motion for the internal liner interfaces are given by
\begin{multline}\label{rliddot}
\ddot{r}_{l,i}=\frac{p_{l,s=i-1}-p_{l,s=i}}{m_{ls}} \cdot 2\pi r_{l,i} \cdot h \\ \left(i=2,3,\ldots,N_{li}-1\right),
\end{multline}
where $p_{l,s}$ is the effective pressure in liner shell $s$.  Each $p_{l,s}$ is comprised of material pressure, $p_{ml,s}$, azimuthal magnetic field pressure, $p_{B_{\theta l},s}$, axial magnetic field pressure, $p_{B_{zl},s}$, and pseudo-pressure due to artificial viscosity, $q_{l,s}$.
The equation of motion for the fuel-liner interface is
\begin{equation}
\label{rgddot}
\ddot{r}_g=\frac{p_{g}+p_{\bar{B}_{zg}}-p_{l,s=1}}{m_{ls}/2}\cdot 2\pi r_g \cdot h,
\end{equation}
where $p_g$ is the gas pressure in the fuel region and $p_{\bar{B}_{zg}}$ is the pressure due to the average axial magnetic field in the fuel region.  The equation of motion for the liner-vacuum interface is
\begin{equation}
\label{rlddot}
\ddot{r}_l=\frac{p_{l,s=N_{ls}}-p_{B_{\theta lv}}-p_{\bar{B}_{zv}}}{m_{ls}/2}\cdot 2\pi r_l \cdot h,
\end{equation}
where $p_{B_{\theta lv}}$ is the pressure of the azimuthal magnetic field evaluated at the liner-vacuum interface ($r=r_l$) and $p_{\bar{B}_{zv}}$ is the pressure due to the average axial magnetic field in the vacuum region.

The magnetic field pressures are simply
\begin{equation}
p_{B_{\mathcal{CR}}}(r)=\frac{B_{\mathcal{CR}}^2(r)}{2\mu_0},
\end{equation}
where $\mathcal{C}$ is the component (either $\theta$ or $z$) and $\mathcal{R}$ is the region (either $g$ for the fuel/gas region, $l$ for the liner region, or $v$ for the vacuum region).  We also note the following. Since we have made an isobaric assumption in the fuel, and since, for MagLIF, the gas pressure typically dominates the magnetic field pressure in the fuel, we simply use the average axial magnetic field in the fuel
\begin{equation}
\bar{B}_{zg}=\frac{\Phi_{zg}}{\pi r_g^2},
\end{equation}
for $p_{\bar{B}_{zg}}$, where $\Phi_{zg}$ is the total axial magnetic flux in the fuel.  By contrast, in the liner region, we assume $B_{zl}(r)/\rho_l(r) = const.$, and thus for $p_{B_{zl},s}$, we use
\begin{equation}
B_{zl,s}(r_{l,s}(t))=\bar{B}_{zl}\frac{\rho_{l,s}}{\bar{\rho}_l},
\end{equation}
where
\begin{equation}
\bar{B}_{zl}=\frac{\Phi_{zl}}{\pi(r_l^2-r_g^2)}
\end{equation}
is the average axial magnetic field in the liner region, $\Phi_{zl}$ is the total axial magnetic flux in the liner region
\begin{equation}
\bar{\rho}_l=\frac{m_l}{\pi(r_l^2-r_g^2)h}
\end{equation}
is the average mass density in the liner region
\begin{equation}
\rho_{l,s}=\frac{m_{ls}}{V_{l,s}},
\end{equation}
is the mass density of shell $s$, and
\begin{equation}
V_{l,s}=\pi (r_{l,i=s+1}^2 - r_{l,i=s}^2) h
\end{equation}
is the volume of shell $s$. For $p_{B_{\theta l}}$, we evaluate our analytic expression for $B_{\theta l}(r)$ (see Eq.~\ref{Bthetal}) at the center of mass of each liner shell, $r_{l,s}(t)$ (see Eq.~\ref{rls}). For $p_{B_{zv}}$, we use the average axial magnetic field in the vacuum
\begin{equation}
\bar{B}_{zv}=\frac{\Phi_{zv}}{\pi \left(r_{rc}^2-r_l^2\right)},
\end{equation}
where $\Phi_{zv}$ is the total axial magnetic flux in the vacuum region of Fig.~\ref{fig:schematic}.  Finally, for $p_{B_{\theta lv}}$, we use the azimuthal magnetic field evaluated at $r=r_l$
\begin{equation}\label{Bthetavatrl}
B_{\theta lv} \equiv B_{\theta l}(r_l) = B_{\theta v}(r_l)=\frac{\mu_0I_l}{2\pi r_l},
\end{equation}
and thus the resulting azimuthal field pressure at $r$=$r_l$ is
\begin{equation}
p_{B_{\theta lv}}=\frac{B_{\theta lv}^2}{2\mu_0}\bigg|_{r=r_l}=\frac{\mu_0I_l^2}{8\pi^2r_l^2}.
\end{equation}

With the magnetic field pressures and liner interface equations of motion thus defined, we can revisit Eqs.~\ref{PEM}--\ref{Pohmapprox}.  Specifically, we can now obtain a more precise expression for the kinetic power
\begin{equation}\label{Pkinexact}
P_{kin}=\sum_i{\delta (p_{B_\theta})_i \cdot \dot{r}_{l,i} \cdot 2\pi r_{l,i} \cdot h},
\end{equation}
where $\delta (p_{B_\theta})_i$ is defined as the difference in azimuthal magnetic field pressure on either side of interface $i$.  This expression describes the rate at which the driving $B_{\theta}$ field does work on each interface, $i$, and thus the rate at which the $B_{\theta}$ field does work on the fuel and liner materials (via material compression and acceleration), as well as on the axial magnetic field, $B_{z}$ (via flux compression). This expression for $P_{kin}$ can be used to replace the approximate expression given in Eq.~\ref{Pkinapprox}, and then from Eq.~\ref{Pohmexact}, we can obtain a more exact expression for $P_{\Omega}$, thus replacing Eq.~\ref{Pohmapprox}. However, in practice, both Eqs.~\ref{Pkinapprox} and \ref{Pkinexact} work fine, since typically, they are within a few percent of each other over most of the implosion.

\subsection{\label{sec:circuit}Liner equation of state, energetics, and ionization}

To model the internal material pressure of the liner shells, we use the expression
\begin{equation}\label{eos}
p_{ml,s}=p_0(\rho_{l,s})+\frac{2}{3}\frac{E_{l,s}}{V_{l,s}}.
\end{equation}
The first term, $p_0(\rho_{l,s})$, is the zero temperature ``cold curve" for the liner material, which is a function of only the mass density, $\rho_{l,s}$.  To model the cold curves of various liner materials, we use the functional form of Birch-Murnaghan\cite{Birch_PR_1947,Murnaghan_AJM_1937,Murnaghan_PNAS_1944}
\begin{multline}\label{p0}
p_0=\frac{3\mathcal{A}_1}{2}\left[\left(\frac{\rho_{l,s}}{\rho_{l0}}\right)^{\gamma_1} - \left(\frac{\rho_{l,s}}{\rho_{l0}}\right)^{\gamma_2}  \right] \\ \times \left\{ 1+\frac{3}{4}\left[\mathcal{A}_2-4\right]\left[\left(\frac{\rho_{l,s}}{\rho_{l0}}\right)^{\sfrac{2}{3}}-1\right] \right\},
\end{multline}
where $\rho_{l0}$ is the liner material's mass density at zero temperature and zero pressure, and where $\mathcal{A}_1$, $\mathcal{A}_2$, $\gamma_1$, and $\gamma_2$ are fitting parameters.  For lithium (Li), beryllium (Be), and aluminum (Al), we find reasonably good fits to SESAME equation of state data\cite{Holian_LANL_1984} using the values provided in Table~\ref{fits}.
\begin{table}
\caption{\label{fits}Nuclear charge, atomic mass number, and cold curve fitting parameters for Eq.~\ref{p0} for various liner materials.}
\begin{ruledtabular}
\begin{tabular}{cccccccc}
 Material & $Z_{nuc}$ & $A$ & $\rho_{l0}$ [kg/m$^3$] & $\mathcal{A}_1$ & $\mathcal{A}_2$ & $\gamma_1$ & $\gamma_2$\\
\hline
Li & 3 & 6.94 & 534 & $11\times 10^9$ & 3.999 & 1.9 & 1.18 \\
Be & 4 & 9.012 & 1845 & $130\times 10^9$ & 3.9993 & 1.85 & 1.18 \\
Al & 13 & 26.98 & 2700 & $76\times 10^9$ & 3.9 & 7/3 & 5/3 \\
\end{tabular}
\end{ruledtabular}
\end{table}

The second term on the right hand side of Eq.~\ref{eos} is the ideal gas thermal contribution to the equation of state, where $E_{l,s}$ is the total thermal energy of the liner material in shell $s$. With finite liner temperature, $E_{l,s}$ can change due to adiabatic compression and/or expansion. Additionally, $E_{l,s}$ can increase because of artificial viscosity and ohmic heating. We have found that including artificial viscosity is useful because it dampens the spring-like reverberations that occur in the liner's wall thickness throughout the implosion. The artificial viscosity model that we use is similar to that discussed in Ref.~\onlinecite{Richtmyer_and_Morton_1967}. Our pseudo-viscous pressure for shell $s$ is given by
\begin{equation}\label{ql}
q_{l,s} =
\begin{cases}
a_q^2\rho_{l,s}(\dot{r}_{l,i=s+1}-\dot{r}_{l,i=s})^2 & \text{for }\dot{r}_{l,i=s}>\dot{r}_{l,i=s+1}\\
0 & \text{for }\dot{r}_{l,i=s}\leq\dot{r}_{l,i=s+1},
\end{cases}
\end{equation}
where $a_q$ is an arbitrary coefficient of $\mathcal{O}(1)$ that specifies the length scale of the artificial viscosity in multiples of the radial thickness of shell $s$
\begin{equation}
\delta_{l,s}= r_{l,i=s+1}-r_{l,i=s}.
\end{equation}
The overall fusion calculations are not very sensitive to the exact value chosen for $a_q$. We find that $a_q$ in the range of 1--3 works well for our purposes. Note Eq.~\ref{ql} states that the artificial viscosity contributes to the shell pressure only during times of compression.  The rate of change of the internal thermal energy of liner shell $s$ is thus given by
\begin{multline}\label{Elsdot}
\dot{E}_{l,s}=-\left( \frac{2}{3}\frac{E_{l,s}}{V_{l,s}} + q_{l,s} \right)\dot{V}_{l,s} \\ + \left( P_r + P_{c} + P_{\Omega} - P_{BB} \right)/N_{ls} \\ \left(s=1,2,\ldots,N_{ls}\right),
\end{multline}
where
\begin{equation}
\dot{V}_{l,s}=2\pi h \left(r_{l,i=s+1} \dot{r}_{l,i=s+1}-r_{l,i=s} \dot{r}_{l,i=s}\right).
\end{equation}
The first term on the right hand side of Eq.~\ref{Elsdot} is due to adiabatic compression/expansion, while the second term is due to artificial viscous heating during compression only. The quantities $P_r$ and $P_{c}$ represent the radiative and thermal conduction loss powers from the fuel region, respectively; they will be described in more detail below.  For simplicity, the powers $P_r$, $P_{c}$, and $P_{\Omega}$ are all assumed to be fully absorbed by the liner and evenly distributed throughout the liner, hence they are each divided by $N_{ls}$ in Eq.~\ref{Elsdot}.\cite{transnote}  The liner is permitted to cool through adiabatic expansion as well as blackbody radiation.  We use blackbody radiation because the liner is optically thick for typical liner materials (Li, Be, Al), wall thicknesses ($>$100~$\mu$m), and average liner temperatures ($<$100~eV).  The blackbody power is given by
\begin{equation}\label{BB}
P_{BB}=\sigma \bar{T}_l^4 \cdot 2\pi r_l \cdot h,
\end{equation}
where $\sigma=5.67\times 10^{-8}$ W/(m$^{2}$$\cdot$K$^{4}$) is the Stefan--Boltzmann constant. For simplicity, energy losses from the liner due to $P_{BB}$ are evenly distributed throughout the liner, hence $P_{BB}$ is also divided by $N_{ls}$ in Eq.~\ref{Elsdot}.  Note that in Eq.~\ref{BB}, we use the average liner temperature, which is given by
\begin{equation}
\bar{T}_l = \frac{2}{3}\frac{E_l}{N_l k} = \frac{2}{3}\frac{\sum_s{E_{l,s}}}{N_{at}\left( 1+\bar{Z}_l \right)k} = \frac{2}{3}\frac{uA\sum_s{E_{l,s}}}{m_l\left( 1+\bar{Z}_l \right)k},
\end{equation}
where $k=1.38\times 10^{-23}$ J/K is the Boltzmann constant, $E_l$ is the total thermal energy of the liner, $N_l=N_{at}(1+\bar{Z}_l)$ is the total number of particles in the liner (ions and electrons), $N_{at}=m_l/(uA)$ is the total number of atomic nuclei in the liner, $u = 1.66\times 10^{-27}$ kg is the unified atomic mass unit, $A$ is the atomic mass number of the liner material (see Table~\ref{fits}), and $\bar{Z}_l$ is the average ionization state of the liner.  We approximate the average ionization state using\cite{Drake_2006}
\begin{equation}
\bar{Z}_l=\min{\left(20\sqrt{\bar{T}_{l,keV}}, Z_{nuc}\right)},
\end{equation}
where $Z_{nuc}$ is the atomic number (nuclear charge) of the atoms comprising the liner material (see Table~\ref{fits}). The average liner temperature in keV is simply
\begin{equation}
\bar{T}_{l,keV} = \frac{k\bar{T}_l}{q_e}\times 10^{-3},
\end{equation}
where $q_e = 1.6\times 10^{-19}$ C (or J/eV) is the charge of an electron.

For reasonable fusion calculations, we have found it necessary to discretize the liner region into concentric thin shells, as described above, because even a compressible cylindrical slab-like model for the liner region (i.e., no internal interfaces, just the fuel-liner and liner-vacuum interfaces connected by the equation of state described above) results in very optimistic fusion yields due to very optimistic fuel compression.  This occurs because in the cylindrical slab-like model, too much mass is assigned to the fuel-liner interface.  The only way to reduce the mass at the fuel-liner interface is to use a finer discretization.  We find that solutions tend to converge when approximately 20 liner shells are used. 

\subsection{\label{sec:circuit}Fuel ionization, energetics, and adiabatic heating}

Within the fuel region, we assume adiabatic compression and a radially constant pressure profile (an isobaric assumption). In reality, there will be small radially dependent pressure waves that reverberate within the fuel; however, the slow liner implosion, which on average is subsonic relative to the fuel's sound speed,\cite{Slutz_PoP_2010} implies that our adiabatic approximation is well justified. This also allows us to assume equal ion and electron temperatures, which we do throughout this semi-analytic model.

To model the energetics in the fuel region, we assume that the fuel is an ideal gas, and thus the isobaric particle pressure is related to the total thermal energy of the fuel, $E_{thg}$, by
\begin{equation}\label{Pg}
p_{g}=\frac{2}{3}\frac{E_{thg}}{V_g},
\end{equation}
where
\begin{equation}
\label{Vg} V_g=2\pi r_g^2h
\end{equation}
is the volume of the gas.  Furthermore, the sum of the thermal energy and the ionization energy gives the total energy of the fuel
\begin{equation}
E_g=E_{thg}+E_{iong}.
\end{equation}
We have found that accounting for the ionization energy is important for MagLIF in cases where the available preheat energy is low ($\sim 100$ J).  For example, about 50 J would be required to fully ionize the deuterium fuel used in recent experiments at the Z facility.\cite{Gomez_PRL_2014} This is a substantial fraction of the 100--300 J of preheat energy that is thought to have coupled to the fuel in these experiments.\cite{Sefkow_PoP_2014}

For the ionization energy of particle species $s$ in the fuel, we have found good fits to tabulated data\cite{CRC_2003} using the analytic expression
\begin{align}\label{Eions}
E_{ion,s} &= 13.6 \cdot q_e \cdot \frac{Z_{nuc,s}^\zeta}{\left(Z_{nuc,s} - \bar{Z}_s + 1\right)^{\nicefrac{1}{\zeta}}} \\
\zeta &= 2.407,
\end{align}
where $Z_{nuc,s}$ is the atomic number (nuclear charge) of atomic species $s$, and $\bar{Z}_s$ is the average ionization state of species $s$.  Here we again make use of the approximate formula\cite{Drake_2006}
\begin{equation}
\bar{Z}_s = \min{\left( 20\sqrt{\bar{T}_{g,keV}}, Z_{nuc,s}\right)},
\end{equation}
where
\begin{align}
\bar{T}_{g,keV} &= \frac{k\bar{T}_g}{q_e}\times 10^{-3} \\
\bar{T}_g &= \frac{2}{3}\cdot \frac{E_{thg}}{\left(N_i + N_e\right) k} \\
N_i &= \sum_s N_{s} \\
N_e &= \bar{Z}_g N_i \\
\bar{Z}_g &= \frac{1}{N_i} \sum_s \bar{Z}_s N_{s},
\end{align}
and where $\bar{T}_{g,keV}$ and $\bar{T}_g$ are the mean temperature of the entire fuel region in units of keV and K, respectively; $N_i$ and $N_e$ are the total number of ions and electrons in the fuel, respectively; $N_{s}$ is the total number of ions of atomic species $s$ in the fuel; and $\bar{Z}_g$ is the average ionization state of the fuel. Finally, summing over all species gives
\begin{equation}
E_{iong} = \sum_s{E_{ion,s} \cdot N_s}.
\end{equation}

The purpose of accounting for various particle species, $s$, is so that we can study the effects of prescribed levels of dopants and/or contaminants (``mix") in the fuel region. Also note that since ionization depends on temperature and temperature depends on ionization, this would require an iterative solution within our overall ODE system solve.  However, since we are only roughly estimating the ionization, and since it changes relatively slowly, we avoid having to implement an iterative solution by holding the ionization fixed during a system solve and updating it only between successive system solves (which is typically every $\sim 100$ ps).

With ionization established, we note the following simple relationships, which will be used throughout the remainder of this manuscript:
\begin{align}
m_g &= \sum_s {uA_s N_s} \nonumber \\ 
&= m_d N_d + m_t N_t + \sum_{s\neq d,t}{uA_s N_s} \\
\bar{m}_i &= m_g/N_i \\
\bar{\rho}_g &= m_g/V_g \\
\bar{n}_i &= N_i/V_g \\
\bar{n}_e &= N_e/V_g \\
\bar{n}_s &= N_s/V_g,
\end{align}
where $m_g$ is the total mass of the fuel; $A_s$ is the atomic mass number of species $s$; $m_d=3.34\times 10^{-27}$ kg is the mass of a deuteron; $m_t=5.01\times 10^{-27}$ kg is the mass of a triton; $N_d$ and $N_t$ are the number of deuterons and tritons in the fuel, respectively; $\bar{m}_i$ is the average mass of an ion in the fuel; $\bar{\rho}_g$ is the average mass density in the fuel; $\bar{n}_i$ and $\bar{n}_e$ are the average ion and electron densities in the fuel, respectively; and $\bar{n}_s$ is the average ion density in the fuel for species $s$.

The dynamics of the total fuel energy are described by
\begin{equation}\label{Egdot}
\dot{E}_g= P_{pdV} + P_{ph} + P_{\alpha}  - P_r - P_{c} - \dot{E}_{ends}
\end{equation}
where $P_{pdV}$ is the adiabatic heating rate, $P_{ph}$ is the fuel preheating rate, $P_\alpha$ is the heating rate due to $\alpha$-particle energy deposition, $P_r$ is the radiative cooling rate, $P_{c}$ is the cooling rate due to electron and ion thermal conduction, and $\dot{E}_{ends}$ is the energy loss rate due to fuel escaping out of the top and/or bottom ends of the imploding cylinder.  The adiabatic heating rate is given by
\begin{equation}
P_{pdV}=p_g\dot{V}_g=\frac{4}{3}E_{thg}\dot{r}_g/r_g.
\end{equation}

\subsection{\label{sec:preheat}Fuel preheating (optionally via laser absorption)}

The fuel preheating rate can be described simply using a square pulse
\begin{equation}\label{Pl}
P_{ph} =
\begin{cases}
0 & \text{for}\quad t<t_{ph}\\
\sfrac{E_{ph}}{\tau_{ph}} & \text{for}\quad t_{ph}\leq t\leq (t_{ph}+\tau_{ph})\\
0 & \text{for}\quad t>(t_{ph}+\tau_{ph}),
\end{cases}
\end{equation}
where $E_{ph}$, $\tau_{ph}$, and $t_{ph}$ are the preheating energy, pulse length, and turn-on time, respectively.

The experimental MagLIF program at Sandia National Laboratories has been investigating the approach of preheating the fuel using the Z beamlet laser.\cite{Gomez_PRL_2014}  In general, MagLIF does not require that the preheating be done with a laser.  One could imagine finding a way to preheat the fuel using some of the energy supplied by the pulsed-power driver, which may enable more total preheat energy, as well as more efficient preheating of the fuel.  Nevertheless, should the preheating be done with a laser, we can account for the beam propagation and deposition analytically.

An analytic description of laser propagation and deposition is particularly useful for quickly evaluating the various MagLIF target designs presently being considered for use at the Z facility.  For example, MagLIF targets (see Fig.~2 in Ref.~\onlinecite{Gomez_PRL_2014}) have consisted of a cylindrical laser entrance channel (LEC), filled will fuel, that resides axially between the bottom of the laser entrance window (LEW) and the top of the imploding region ($z=h$ in Fig.~\ref{fig:schematic}).  We define the axial length of this channel as $\Delta z_{LEC}$.  Furthermore, a beam dump has been used, residing below the imploding region (below $z=0$ in Fig.~\ref{fig:schematic}).  This beam dump is simply a reservoir of fuel that is present in case the laser propagates completely through the imploding region. If the fuel reservoir were not present, the laser could hit and ablate high-Z material from various target and/or electrode surfaces, which could spray back into the fuel of the imploding region, providing a source of contaminant mix.

To account for the energy absorbed in each region (i.e., the LEC region, the imploding region, and the beam dump region), we use the inverse bremsstrahlung model outlined in Ref.~\onlinecite{Slutz_PoP_2010}. By assuming an ionization state for the fuel while it is interacting with the beam; for example, we use the fully-ionized condition given by
\begin{equation}
\bar{Z}_{b} = \frac{1}{N_i} \sum_s Z_{nuc,s} N_{s};
\end{equation}
and by ignoring thermal conduction and hydrodynamic motion (i.e., we assume fast isochoric laser heating), the heating of the fuel by laser light can be described by\cite{Z&R_2002}
\begin{align}
\label{dIdz}   \frac{d\varepsilon_{b}}{dt} &= \frac{dI_{b}}{dz} = -\kappa(z,t)I_{b} \\
\label{kzt} \kappa(z,t) &= \frac{\nu_{ei}}{c}\frac{\omega_p^2}{\omega_b^2}\left[1-\frac{\omega_p^2}{\omega_b^2}\right]^{-\frac{1}{2}},
\end{align}
where $\varepsilon_{b} (z,t)$ is the energy density in the fuel that is interacting with the beam, $I_{b}(z,t)$ is the beam intensity in the fuel, $\kappa (z,t)$ is the absorption coefficient, $\nu_{ei}$ is the electron-ion collision frequency, $c=3\times 10^8$ m/s is the speed of light in vacuum, $\omega_p$ is the plasma frequency in the fuel, and $\omega_b$ is the laser frequency.  The plasma frequency is given by\cite{Chen_1984}
\begin{equation}\label{wp}
\omega_p = \sqrt{\frac{\bar{n}_e q_e^2}{m_e \epsilon_0}},
\end{equation}
where $m_e=9.1\times 10^{-31}$ kg is the mass of an electron, and $\epsilon_0=1/(\mu_0 c^2)$ is the permittivity of free space. The laser frequency is given by
\begin{equation}
\omega_b=2\pi c / \lambda_b,
\end{equation}
where $\lambda_b$ is the laser wavelength.  For preheating the fuel with the Z beamlet laser\cite{Rambo_AO_2005} at the Z facility, $\lambda_b=532$ nm.
The electron-ion collision frequency is given by\cite{Callen}
\begin{equation}
\label{nueizblabs}\nu_{ei} = \frac{4\sqrt{2\pi}\sum_s{\left(\bar{n}_s Z_{nuc,s}^2\right)} q_e^4 \ln\Lambda}{\left\{4\pi\epsilon_0\right\}^2 3\sqrt{m_e}\left(kT_b\right)^{\nicefrac{3}{2}}},
\end{equation}
where we have accounted for multiple ions species, $s$; $T_b$ is the temperature of the fuel that is interacting with the beam; and $\ln\Lambda$ is the Coulomb logarithm (see Appendix~\ref{loglambda}).  Next, using
\begin{equation}
\label{kTb} kT_{b} = \frac{2}{3}\cdot\frac{\varepsilon_b(z,t)}{\bar{n}_i+\bar{n}_e},
\end{equation}
we factor out the $\varepsilon_b(z,t)$ dependence from Eq.~\ref{kzt} and evaluate the remaining factors at $t=t_{ph}$; we define the result as
\begin{equation}
\label{kappatph} \tilde{\kappa}(t_{ph})\equiv  \left[ \kappa(z,t) \cdot \varepsilon_{b}^{\nicefrac{3}{2}}(z,t) \right]_{t=t_{ph}}.
\end{equation}
This implies that the only $\varepsilon(z,t)$ dependence in $\kappa(z,t)$ is that shown explicitly by substituting Eq.~\ref{kTb} into Eq.~\ref{nueizblabs} and then Eq.~\ref{nueizblabs} into Eq.~\ref{kzt}. However, there is also an implicit $\varepsilon(z,t)$ dependence in the calculation of $\ln\Lambda$ in Eq.~\ref{nueizblabs}, which we neglect for simplicity.  For the $kT$ dependence in $\ln\Lambda$, we simply use the average energy per particle required to meet our full ionization assumption.  That is, we use Eq.~\ref{Eions} with $\bar{Z}_s = Z_{nuc,s}$, and thus we use $kT=13.6\cdot q_e \cdot \sum_s{Z_{nuc,s}^\zeta}$ in the $\ln\Lambda$ calculation of Eq.~\ref{nueizblabs}.

The purpose of factoring out $\varepsilon_{b}(z,t)$ in Eqs.~\ref{dIdz}--\ref{kappatph} is to make it clear that, for this subroutine of calculating laser absorption and propagation, we are simply freezing all dynamic variables [other than $\varepsilon_{b}(z,t)$ and $I_{b}(z,t)$] to their values at the time when the laser is first applied (i.e., our assumption of fast isochoric laser heating). With this done, Eq.~\ref{dIdz} becomes
\begin{equation}
\frac{d\varepsilon_{b}}{dt}=\frac{dI_{b}}{dz}=-\tilde{\kappa}(t_{ph})\cdot \varepsilon_{b}^{-\nicefrac{3}{2}}(z,t)\cdot I_{b}(z,t),
\end{equation}
which has an exact solution given by\cite{Slutz_PoP_2010} 
\begin{align}
\varepsilon_{b}(z,t) &= \varepsilon_{b}(z_{LEW},t)\left[1-\frac{z_{LEW}-z}{z_{LEW}-z_f(t)}\right]^{\nicefrac{2}{3}} \\
I_{b}(z,t) &= I_{b}(z_{LEW},t)\left[1-\frac{z_{LEW}-z}{z_{LEW}-z_f(t)}\right]^{\nicefrac{2}{3}},
\end{align}
where
\begin{align}
\varepsilon_{b}(z_{LEW},t) &= \left[\frac{5}{2}\cdot \tilde{\kappa}(t_{ph})\cdot I_{b}(z_{LEW},t) \cdot (t-t_{ph}) \right]^{\nicefrac{2}{5}} \\
z_f(t) &= z_{LEW} - \frac{5}{3}\cdot \frac{I_{b}(z_{LEW},t)}{\varepsilon_{b}(z_{LEW},t)}\cdot (t-t_{ph}),
\end{align}
and where $z_{LEW}=h+\Delta z_{LEC}$ is the axial location of the LEW (where the beam first interacts with the fuel), and $z_f(t)$ is the axial location of the laser heating front as the beam ``bleaches" its way through the fuel.  Note that we have written this solution for a downward propagating beam, i.e., in the $-\hat{z}$ direction, and that this solution is applicable only for $t_{ph}\leq t \leq (t_{ph}+\tau_{ph})$ and $z_f(t) \leq z \leq z_{LEW}$.

The energy deposition rates in the LEC region, in the imploding/fusion region, and in the beam dump region, are given by
\begin{align}
P_{LEC} &= \pi r_b^2 \cdot \left[ I_b(z_{LEW},t) - I_b(h,t) \right] \\
P_{ph} &= \pi r_b^2 \cdot \left[ I_b(h,t) - I_b(0,t) \right] \label{Pph} \\
P_{dump} &= \pi r_b^2 \cdot I_b(0,t),
\end{align}
respectively, where $r_b$ is the radius of the beam and $I_b(z_{LEW},t)=P_{in}/(\pi r_b^2)$.  Here, $P_{in}$ is the laser power that makes it through the LEW and enters the fuel.

For fusion calculations, we only need the energy deposited in the imploding region from $z=0$ to $h$, hence we only need $P_{ph}$ as given by Eq.~\ref{Pph} (or by Eq.~\ref{Pl} if ignoring laser absorption).  We assume that the energy deposited in the imploding region is instantly distributed uniformly in both the axial and radial directions. In the axial direction, this is a reasonable assumption since thermal conduction is uninhibited in this direction (i.e., the applied magnetic field that undergoes flux compression, $B_{zg}$, is axially aligned, and thus inhibits thermal transport only in the radial direction).  Furthermore, fully-integrated 2D radiation magnetohydrodynamics simulations have shown that the preheat energy redistributes axially on a timescale that is fast compared to the implosion time.\cite{Sefkow_PoP_2014}  In the radial direction, full radiation magnetohydrodynamics simulations show that a pressure wave (a ``blast wave") is generated by the rapid preheating of the fuel, which expands radially outward from approximately the beam radius until it hits and reflects off of the liner's inner surface.\cite{Sefkow_PoP_2014} Subsequent pressure waves then reverberate within the fuel, but settle to a reasonably isobaric state on a timescale that is short relative to the overall implosion time.

Note that even though the pressure (overall energy density) in the fuel is assumed to be distributed uniformly in both the axial and radial directions, the fuel temperature and density profiles are assumed to be uniform only in the axial direction.  Moreover, the radial temperature and density profiles are significantly affected by the radial extent of the fuel that is initially preheated relative to the radial extent of the overall fuel region.  This is discussed in more detail below.

For the radius of the beam, $r_b$, we ignore laser-plasma interactions and use the beam radius at the axial midpoint of the imploding region.  Furthermore, we assume that the beam follows the focusing and defocusing cones described by\cite{Geissel_personal_waist_2014}
\begin{equation}\label{beameq}
r_b = \frac{1}{2} \left[\frac{\Delta z_b}{f/\#}+\frac{\diameter_{bf}}{1+c_b\Delta z_b}\right],
\end{equation}
where $\Delta z_b$ is the distance from the beam's focal plane to the axial position of interest, $f/\#$ is the beam's ``$f$-number'', $\diameter_{bf}$ is the diameter of the beam in the beam's plane of best focus, and $c_b=428.6$ m$^{-1}$ is a fitting parameter. Note that the second term in Eq.~\ref{beameq} is significant only in regions close to the focal plane; it is a first-order corrective term for describing the ``waist'' of the beam.  As an example of using Eq.~\ref{beameq}, we consider the experiments of Ref.~\onlinecite{Gomez_PRL_2014}, where the $f$/10 Z beamlet laser was focused to a 250-$\mu$m spot size, roughly 3.5 mm above the LEW; thus, with $\Delta z_b=3.5$ mm, Eq.~\ref{beameq} gives a spot size on the LEW of 450 $\mu$m, as quoted in Ref.~\onlinecite{Gomez_PRL_2014}.  Additionally, the length of the LEC was about 2.1 mm, and the length of the imploding region was about 7.5 mm; thus, with $\Delta z_b=\left(3.5 + 2.1 + 7.5/2\right) = 9.35$~mm, Eq.~\ref{beameq} gives $r_b = 492$ $\mu$m.  For reference, $r_{g0}$ was 2.325 mm.

Note that we have not accounted for absorption or scattering due to the laser entrance window.  For this, we rely on separate laser-only experiments that measured transmission through foils similar to those used for LEWs on MagLIF experiments; this was done for various spot sizes on the foil and will be presented in a future publication.\cite{Geissel_personal_2014,Geissel_DPP_2014}  For now, we simply note that
\begin{equation}\label{Pin}
P_{in} = \mathcal{T}_{LEW} \cdot P_{laser},
\end{equation}
where $\mathcal{T}_{LEW}$ is the transmission through the LEW and $P_{laser}$ is the laser power on the vacuum side of the LEW.  We comment that modeling, simulating, and experimentally diagnosing laser transmission, propagation, and absorption in such a system is a very challenging problem, even for much more sophisticated simulation codes; these are presently very active areas of research within the MagLIF program and elsewhere.

\subsection{\label{sec:circuit}Fuel heating via magnetized $\alpha$-particle energy deposition}

The next term to calculate in Eq.~\ref{Egdot} is the heating rate due to $\alpha$-particle energy deposition, $P_\alpha$.  To do so, we note that the $\alpha$ particles are produced only by DT reactions and that the energy carried by each $\alpha$ particle is
\begin{equation}
Q_{\alpha}=3.5\times 10^6\cdot q_e.
\end{equation}
Following Ref.~\onlinecite{Basko_NF_2000}, the fraction of $Q_\alpha$ that is deposited in our magnetized cylindrical fuel is calculated using
\begin{equation}
 f_\alpha = \frac{x_\alpha+x_\alpha^2}{1+13x_\alpha/9+x_\alpha^2}, \label{eqn:fa}
\end{equation}
where
\begin{align}
x_\alpha &= \frac{8}{3}\left(\frac{r_g}{l_\alpha}+\frac{b^2}{\sqrt{9b^2+1000}}\right) \label{eqn:xa}\\
l_\alpha &= \left\{4\pi\epsilon_0\right\}^2 \cdot \frac{3}{4\sqrt{2\pi}} \cdot \frac{m_\alpha v_{\alpha 0} \left(kT_g\right)^{\nicefrac{3}{2}}}{\bar{n}_e Z_{\alpha}^2 q_e^4 m_e^{1/2} \ln\Lambda} \label{lalpha}\\
b &= \frac{r_g}{r_{\alpha L}} \\
r_{\alpha L} &= \frac{m_\alpha v_{\alpha 0}}{Z_\alpha q_e \bar{B}_{zg}},
\end{align}
and where $l_\alpha$ is the mean free path of an $\alpha$ particle, $m_\alpha=6.64\times 10^{-27}$ is the mass of an $\alpha$ particle, $v_{\alpha 0}=\sqrt{2Q_\alpha/m_\alpha}$ is the birth velocity of an $\alpha$ particle, $Z_\alpha=2$ is the charge of an $\alpha$ particle, $r_{\alpha L}$ is the Larmor radius of an $\alpha$ particle, and the $\ln\Lambda$ calculation is described in Appendix~\ref{loglambda}. The $\alpha$ particle heating rate is then given by
\begin{equation}
P_{\alpha}=\dot{N}_{dt} Q_\alpha f_\alpha,
\end{equation}
where $\dot{N}_{dt}$ is the primary DT reaction rate, which is given below (see Eq.~\ref{Ndtdot}).

\subsection{\label{sec:circuit}Model of fuel hot spot and dense outer shelf regions}

For the remaining terms in Eq.~\ref{Egdot}, which all describe energy losses from the fuel, and for describing magnetic flux loss due to the Nernst thermoelectric effect, we need to assume something about the temperature and density gradients in the fuel.  For this semi-analytic model of MagLIF, we assume temperature and density profiles that are comprised of two parts: a low-density hot spot region and a cold dense shelf region (see Fig.~\ref{fig:schematic}). This separation into two regions occurs when the preheating is done rapidly across a region from $r=0$ to $r_{ph0}$, where $r_{ph0} <r_g(t_{ph})$.\cite{unote}  This separation occurs because the rapid preheating causes a blast wave to propagate radially outward from $r_{ph0}$ toward the outer radius of the overall fuel region, $r_g(t)$.  This blast wave redistributes mass by boring out a low-density hot spot and pushing colder non-preheated fuel up against the liner's inner surface.

For the low density hot spot region, we assume a fuel (gas) temperature profile of the form
\begin{multline}
\label{Tgr}T_g (r) = T_c \left\{1-\left(\frac{r}{r_h}\right)^\xi\left[1-\left(\frac{T_B}{T_c}\right)\right]\right\}  \\
\text{for }0\leq r\leq r_h \leq r_g,
\end{multline}
and a radiation temperature given by
\begin{equation}
T_r(r) = const. = T_B \quad\text{for }0\leq r\leq r_h \leq r_g,
\end{equation}
where $T_c$ is the central (peak) temperature at $r=0$, $r_h$~is the outer radius of the hot spot region (see Fig.~\ref{fig:schematic}), $\xi$ is the power-law parameter that specifies the curvature of the profile, and $T_B$ is the brightness temperature, which is found iteratively while calculating the radiative cooling rate, $P_r$ (see discussion about $P_r$ below, in Sec.~\ref{sec:radlosses}).  By comparing to detailed calculations (see for example Fig.~5 in Ref.~\onlinecite{Slutz_PoP_2010}), we find the hot spot profile to be reasonably well modeled using $\xi\approx 2$ with Nernst effects, and $\xi\approx 3.5$ without Nernst effects.  The results are not overly sensitive to this choice.  We have found good agreement using anything from $\xi= 2$ to $\xi= 8$.

Because of our isobaric assumption, specifying the temperature or the pressure at any point in the fuel automatically determines the other, i.e., $\rho_g(r)\cdot T_g(r)={const}$. Thus, Eq.~\ref{Tgr} means that the density profile in the hot spot region is given by
\begin{multline}
\label{rhogr}\rho_g (r) = \rho_c \left\{1-\left(\frac{r}{r_h}\right)^\xi\left[1-\left(\frac{T_B}{T_c}\right)\right]\right\}^{-1} \\
\text{for }0\leq r\leq r_h \leq r_g,
\end{multline}
where $\rho_c = \bar{\rho}_g \bar{T}_g/T_c$ is the density at $r=0$.

In the cold dense shelf region, the fuel temperature and the radiation temperature are given by
\begin{equation}\label{Tgrshelf}
T_g(r) = T_r(r) = T_B \cdot \left( \frac{r_h}{r} \right)^{\nicefrac{1}{4}} \quad \text{for }r_h \leq r \leq r_g,
\end{equation}
and thus by our isobaric assumption, the fuel density in the shelf region is given by
\begin{equation}\label{rhogrshelf}
\rho_g(r) = \frac{\rho_c T_c}{T_B}\cdot \left( \frac{r}{r_h} \right)^{\nicefrac{1}{4}} \quad \text{for }r_h \leq r \leq r_g.
\end{equation}
The temperature gradients in each region are thus given by
\begin{equation}
\frac{\partial T_g}{\partial r} =
\begin{cases}
-\frac{\xi}{r}   \left(T_c-T_B\right)  \left(\frac{r}{r_h}\right)^\xi   & \text{for}\quad 0<r\leq r_h \leq r_g \\
-\frac{T_B}{4r} \left(\frac{r_h}{r}\right)^{\nicefrac{1}{4}} & \text{for}\quad r_h < r \leq r_g.
\end{cases}
\end{equation}
Note that the temperature and density profiles are nearly flat in the shelf region and that $T_g(r)$, $T_r(r)$, $\nabla T_g$, $\rho_g(r)$, and $\nabla\rho_g$ are all finite throughout the entire fuel region.  The reasons for using these functional forms, as well as how to find $T_c$, $\rho_c$, $r_h$, and $T_B$, are deferred to the discussion below on radiative losses from the fuel (Sec.~\ref{sec:radlosses}).

With these profiles, we can now define the following radially dependent number densities in the fuel:
\begin{align}
n_i &\equiv n_i(r) = N_i \cdot \rho_g(r)/m_g \\
n_e &\equiv n_e(r) = \bar{Z}_g n_i \\
n_s &\equiv n_s(r) = N_s \cdot \rho_g(r)/m_g, \label{eqn:n_s}
\end{align}
which are for the ions, electrons, and the ions of species $s$, respectively. Also, by assuming a ``frozen-in" condition for the axial magnetic field in the fuel plasma, we have $B_{zg}(r)/\rho_g(r) = const. = \bar{B}_{zg}/\bar{\rho}_g$, and thus
\begin{equation}
B_{zg}\equiv B_{zg}(r) = \frac{\bar{B}_{zg}}{\bar{\rho}_g} \rho_g(r).
\end{equation}
Note that the Nernst effect (discussed below) actually breaks the frozen-in condition ({\it cf}. Fig.~5 in Ref.~\onlinecite{Slutz_PoP_2010}).  Nevertheless, this expression for $B_{zg}(r)$ provides us with a simple first-order approximation of the radial dependence of the axial magnetic field that is accurate enough for the purposes of this model.

\subsection{\label{sec:radlosses}Radiative losses}

The next term to calculate in Eq.~\ref{Egdot} is the radiative cooling rate, $P_r$.  This calculation determines the fuel temperature and density profiles described in Eqs.~\ref{Tgr}--\ref{rhogrshelf} by finding $T_B$ iteratively while applying isobaric and conservation of mass arguments.  The formulation is derived from a two-temperature gray model of nonequilibrium radiative diffusion.  For more details on this type of model, see the discussion on pages 261--262 in Ref.~\onlinecite{Drake_2006}.

Using a two-temperature model was motivated after studying the results of full radiation magnetohydrodynamics simulations of MagLIF and observing the following: (1) the radiation temperature is nearly constant throughout the fuel, dropping only slightly in the shelf region as $r$ approaches $r_g$; (2) the radiation temperature is significantly lower than the fuel temperature in the hot spot region; (3) the radiation temperature is roughly equal to the fuel temperature in the shelf region; (4) the radiative flux across the fuel-liner interface is well approximated by $(1-\alpha_s)\sigma T_s^4$, where $\alpha_s\gtrsim 0.9$ is the albedo of the liner's inner surface and $T_s(r)$ is the temperature in the shelf region (both the fuel and radiation temperatures since they are equal in the shelf region).

These phenomena occur because the liner's inner surface heats and compresses rapidly to a point where it can no longer absorb or transmit radiation efficiently.  Thus, this surface begins to reflect/reemit radiation back into the fuel region, effectively trapping a significant portion of the radiation in the fuel region.  Moreover, thermal diffusion from the liner's hot inner surface to the colder material deeper within the liner region is slow and on a timescale that is long compared to the implosion time. This slow diffusion is due in part to the high heat capacity of the liner material, and it helps to enable the radiation trapping.

These observations inspired the following two-temperature gray model for radiative losses over the volume of the fuel, where opacity effects become stronger in and near the cold dense shelf region.  From Eq.~6.62 in Ref.~\onlinecite{Drake_2006}, we have
\begin{equation}\label{epsdot}
\dot{\varepsilon}_g = 4\kappa_r\sigma T_r^4 - 4\kappa_p\sigma T_g^4,
\end{equation}
where $\varepsilon_g$ is the energy density of the fuel, $\kappa_r$ is an averaged opacity described by Eq.~6.60 in Ref.~\onlinecite{Drake_2006}, and $\kappa_p$ is the Planck mean opacity.  The first term on the right hand side of Eq.~\ref{epsdot} describes the rate at which the fuel absorbs energy from the radiation field, while the second term describes the rate at which the fuel loses energy to the radiation field. Since the radiation is trapped in the fuel region by the liner's hot inner surface, and since, in the shelf region, the fuel mass and radiation fields are in thermal equilibrium, we assume that $\kappa_r \approx \kappa_p$, and thus
\begin{equation}\label{epsgdot}
\dot{\varepsilon}_g = - 4\kappa_p\sigma T_g^4 \left[ 1-\left(\frac{T_r}{T_g}\right)^4\right].
\end{equation}
Note that a consequence of Eq.~\ref{epsgdot} is that there is no net energy exchange between the fuel and the radiation field when the two are in thermodynamic equilibrium ($T_r=T_g$).  Next, from Eq.~5.22 in Ref.~\onlinecite{Z&R_2002}, we have
\begin{equation}\label{kappasigmaT}
4\kappa_p\sigma T_g^4 = J = A_{br}\cdot \bar{Z}_g^2 n_i n_e \sqrt{T_{g}},
\end{equation}
where $J$ is the frequency-integrated emission coefficient for the bremsstrahlung mechanism and $A_{br} = 1.57 \times 10^{-40}$ ${\rm m^3 \cdot K^{-\frac{1}{2}} \cdot J/s}$.\cite{bremsnote}  Substituting Eq.~\ref{kappasigmaT} into Eq.~\ref{epsgdot} and integrating over the volume of the fuel from $r'=0$ to $r'=r$ gives the cumulative radiative cooling power from the fuel as a function of $r$
\begin{equation}\label{Prv}
P_{rv}(r)=A_{br}\cdot 2\pi h \cdot \bar{Z}_g^2\bigintsss_0^{r} n_i n_e \sqrt{T_{g}}\left[1-\left(\frac{T_{r}}{T_{g}}\right)^4\right]r'dr'.
\end{equation}

Note that since we have set $T_r = T_g$ in the shelf region, the contribution of the shelf region to the $P_{rv}$ integral is zero; this is consistent with the fuel's mass and radiation field being in thermodynamic equilibrium in the shelf region.  The first term in Eq.~\ref{Prv} represents a radiation source and the second term represents a radiation sink due to opacity and absorption.  Note that for the case of $T_r=0$ and constant density and temperature profiles, integrating Eq.~\ref{Prv} over the entire fuel region gives the familiar bremsstrahlung radiation power from an optically-thin volume emitter
\begin{equation}
\bar{P}_{rv} = A_{br}\cdot \bar{Z}_g^2 \bar{n}_i \bar{n}_e \sqrt{\bar{T}_g} \cdot \pi r_g^2 \cdot h.
\end{equation}

Next, to find $T_r$, we invoke the additional constraint that, in the shelf region ($r_h\leq r\leq r_g$), the volume radiation described by Eq.~\ref{Prv} must be consistent with the gray body surface radiation described by 
\begin{equation}\label{Prsor}
P_{rs}(r)=(1-\alpha_s)\sigma T_r^4(r) \cdot 2\pi r \cdot h.
\end{equation}
To do this, we first note that as long as $P_{rv}(r_h)=P_{rs}(r_h)$, then $P_{rv}(r)=P_{rs}(r)$ throughout the shelf region because of the functional form assumed for the shelf temperature in Eq.~\ref{Tgrshelf}, i.e., $T_g(r) = T_r(r) \propto r^{-1/4}$.  This functional form was chosen to ensure that Eq.~\ref{Prsor} would return a constant value, and thus enforce our assumption of no radiative loss or gain within the shelf region. We also note that, in the hot spot region ($0<r\leq r_h$), we have assumed $T_r = const. = T_B$, and thus $T_r(r_h)=T_B$.  Therefore, 
to ensure consistency in the shelf region, we must find the values of $r_h$ and $T_B$ that satisfy
\begin{equation}\label{Prh1}
P_{rv}(r_h)= P_{rs}(r_h),
\end{equation}
where
\begin{align}
P_{rv}(r_h) &= A_{br}\cdot 2\pi h \cdot \bar{Z}_g^2\bigintsss_0^{r_h} n_i n_e \sqrt{T_{g}}\left[1-\left(\frac{T_{B}}{T_{g}}\right)^4\right]rdr  \\
P_{rs}(r_h)&= (1-\alpha_s)\sigma T_B^4 \cdot 2\pi r_h \cdot h.\label{Prh2}
\end{align}

To find the values of $r_h$ and $T_B$ that satisfy Eqs.~\ref{Prh1}--\ref{Prh2}, we use a simple bisection method and evaluate the expressions for $P_{rv}(r_h)$ and $P_{rs}(r_h)$ iteratively until $P_{rv}(r_h) = P_{rs}(r_h)$ to within 1\%.\cite{bisection_vs_best_guess} Also, before we can test our iterative guesses for $r_h$ and $T_B$ in Eqs.~\ref{Prh1}--\ref{Prh2}, we need to find $\rho_c$ and $T_c$ to complete the profile definitions given in Eqs.~\ref{Tgr}--\ref{rhogrshelf}.  To do this, we use our isobaric assumption and conservation of mass arguments. Additional computational details for finding $r_h$, $T_B$, $\rho_c$, and $T_c$ are provided in Appendix~\ref{rhnotes}.

Throughout a given simulation, $\alpha_s = 0.9$ is held fixed. However, the results are not very sensitive to this choice.  In practice, we find that any reasonable value, say in the range of 0.5--0.95, works fine.

Since, in the shelf region, $P_{rv}(r)=P_{rs}(r)=const.=P_{rv}(r_h)=P_{rv}(r_g)$, the total radiative cooling rate for Eq.~\ref{Egdot} can be taken from anywhere in the shelf region; we take
\begin{equation}\label{eqPr}
P_r = P_{rv}(r_g).
\end{equation}

Note that in this model, if the entire fuel region is preheated uniformly, then the cold dense shelf region will never exist (only a hot spot region will exist), and thus all of the fuel mass will contribute to the radiation losses.  Also, due to the isobaric assumption and the conservation of fuel mass, the absence of a shelf region means that the central peak temperature, $T_c$, will be lower for the same amount of energy deposited in the fuel.  For these reasons, uniformly preheating the entire fuel is not optimal. Moreover, preheating only a central portion of the fuel, say from $r=0$ to $r_{ph0}\approx 0.5\cdot r_g(t_{ph})$, broadens the optimal MagLIF operating space by enabling the use of higher initial fuel densities (i.e., if these higher initial fuel densities were used in cases where all of the fuel was preheated, then the radiation losses would cool the preheated fuel too rapidly to obtain good fusion yield).  This is fortunate since the experimental MagLIF program at Sandia is presently investigating the approach of preheating the fuel via a laser with a small beam radius relative to $r_g(t_{ph})$.

Finally, we comment that assuming this shelf region is present in experiments may be idealistic.  Azimuthal or other 3D asymmetries (particularly during laser preheating) might make it impossible to establish or maintain this shelf region.  If a shelf region cannot be maintained (which would be difficult to diagnose experimentally), then the optimal operating space for MagLIF might not be as broad as that suggested by simulations where the initial fuel density is scanned over and $r_{ph0}<r_g(t_{ph})$.

\subsection{\label{sec:circuit}Magnetized electron and ion thermal conduction losses}

The next term to calculate in Eq.~\ref{Egdot} is the energy loss rate due to electron and ion thermal conduction.  To do this, we use the Epperlein-Haines transport equations for the electrons\cite{Epperlein-Haines_PoF_1986} and the Braginskii transport equations for the ions.\cite{Braginskii_1965} The energy loss rate due to electron thermal conduction is given as a function of $r$ by
\begin{equation}
\tilde{P}_{ce}(r)=2\pi r h \cdot \kappa_e(x_e) \cdot k\frac{\partial T_{g}}{\partial r},
\end{equation}
where
\begin{equation}
\kappa_e = \frac{n_ekT_g\tau_{ei}}{m_e}\cdot \frac{6.18+4.66x_e}{1.93+2.31x_e+5.35x_e^2+x_e^3}
\end{equation}
is the coefficient for electron thermal conduction perpendicular to $B_{zg}$, $x_e \equiv \omega_{ce} \tau_{ei}$ is the electron Hall parameter, $\omega_{ce}=q_eB_{zg}/m_e$ is the electron cyclotron frequency, and $\tau_{ei}=1/\nu_{ei}$ is the average time between electron-ion collisions.  Here, $\nu_{ei}$ is the electron-ion collision frequency given by\cite{Callen}
\begin{equation}\label{nueic}
\nu_{ei} = \frac{4\sqrt{2\pi}\left(\sum_s{n_s \bar{Z}_s^2}\right) q_e^4 \ln\Lambda}{\left\{4\pi\epsilon_0\right\}^2 3\sqrt{m_e}\left(kT_g\right)^{\nicefrac{3}{2}}},
\end{equation}
where we have accounted for multiple ion species, $s$, and where the $\ln\Lambda$ calculations are described in Appendix~\ref{loglambda}.

Similar to electron thermal conduction, the energy loss rate due to ion thermal conduction is given as a function of $r$ by
\begin{equation}
\tilde{P}_{ci}(r)=2\pi r h \cdot \kappa_i(x_i) \cdot k\frac{\partial T_{g}}{\partial r},
\end{equation}
where
\begin{equation}
\kappa_i = \frac{n_ikT_g\tau_{ii}}{\bar{m}_i}\cdot \frac{2.645+2x_i^2}{0.677+2.70x_i^2+x_i^4}
\end{equation}
is the coefficient for ion thermal conduction perpendicular to $B_{zg}$, $x_i \equiv \omega_{ci} \tau_{ii}$ is the ion Hall parameter, $\omega_{ci}= q_e \bar{Z}_g B_{zg}/\bar{m}_i$ is the ion cyclotron frequency, and $\tau_{ii}=1/\nu_{ii}$ is the average time between ion collisions.  Here, $\nu_{ii}$ is the ion-ion collision frequency given by\cite{Callen}
\begin{equation}\label{nuiic}
\nu_{ii} = \frac{4\sqrt{\pi} n_{h} \bar{Z}_{h}^4 q_e^4 \ln\Lambda}{\left\{4\pi\epsilon_0\right\}^2 3\sqrt{\bar{m}_h}\left(kT_g\right)^{\nicefrac{3}{2}}} \left(1+ \sqrt{2} \sum_{s\neq d,t}{\frac{n_s \bar{Z}_s^2}{n_h \bar{Z}_h^2}}\right),
\end{equation}
where $n_h\equiv n_d + n_t$, $\bar{Z}_h\equiv \bar{Z}_d = \bar{Z}_t$, and where we have assumed that the plasma is comprised mainly of fuel deuterons and tritons (subscript $h$ for hydrogen isotopes), with an average dominant ion mass of
\begin{equation}
\bar{m}_{h}\equiv \frac{N_d m_d + N_t m_t}{N_d + N_t}
\end{equation}
that is much less than the mass~$m_{s\neq d,t}$ of any of the dopant/contaminant particles.  The $\ln\Lambda$ calculations are described in Appendix~\ref{loglambda}.

Finally, the total radially dependent energy loss rate due to thermal conduction is given by
\begin{equation}
\tilde{P}_{c}(r) = \tilde{P}_{ce}(r) + \tilde{P}_{ci}(r).
\end{equation}
For this calculation, we use radially dependent parameters $n_i(r)$, $n_e(r)$, $x_e(r)$, and $x_i(r)$ [and thus $B_{zg}(r)$, $\nu_{ei}(r)$, $\nu_{ii}(r)$, $n_h(r)$, $n_s(r)$, and $\ln\Lambda (r)$], since they are all readily calculable from our given expressions for $T_g(r)$ and $\rho_g(r)$.

While studying thermal conduction using full radiation magnetohydrodynamics simulations (as well as our hot spot model), we observed that ion thermal conduction dominates over electron thermal conduction in the inner regions of the fuel hot spot. By contrast, near the edge of the hot spot (i.e., $r \approx r_h$), the thermal transport is dominated by electron conduction.  Thus there is a handoff from ions to electrons at some radius in the fuel hot spot region.   Because of this, and because our hot spot profile is prescribed and determined by the radiation model, we wish to be conservative with regards to the expected losses due to thermal conduction, and therefore we look for a maximum in $\tilde{P}_{c}(r)$.  That is, for the thermal conduction loss power from the hot spot region to the shelf or liner region, we use
\begin{equation}
P_{ch} = \max \left\{\tilde{P}_c(r)\right\} \quad \text{for }0<r\leq r_h \leq r_g.
\end{equation}
For the thermal conduction loss power from the shelf region to the liner region, we simply use
\begin{align}
P_{cs} &= \tilde{P}_c(r_g).
\end{align}
The overall thermal conduction loss power from the fuel to the liner depends on whether or not the shelf region exists, and thus, for the thermal conduction power in Eq.~\ref{Egdot}, we use
\begin{equation}
P_{c} = 
\begin{cases}
P_{cs} & \text{for }r_h<r_g\\
P_{ch} & \text{for }r_h=r_g.
\end{cases}
\end{equation}
Note that even if a shelf region exists, we still need to calculate $P_{ch}$ in order to estimate the shelf erosion rate (see Eq.~\ref{mdot} below).

\subsection{\label{sec:endlosses}End losses}

The final term to calculate in Eq.~\ref{Egdot} is the energy loss rate due to fuel losses out of the ends of the imploding cylinder, $\dot{E}_{ends}$.  We estimate this loss by assuming that the fuel (with its associated mass and energy densities) flows across the top ($z=h$) and bottom ($z=0$) planes of the imploding region at the hydrodynamic sound speed, which is determined by the pressure and mass density of the fuel.  Therefore, the energy losses out of the top and bottom planes, respectively, are given by
\begin{align}\label{endlosses}
\dot{E}_{top} &= (3/4)^4 \cdot \frac{E_g}{V_g} \int_0^{r_{top}(t)} c_g(r) \cdot 2\pi r \cdot dr \\
\dot{E}_{bot} &= (3/4)^4 \cdot \frac{E_g}{V_g} \int_0^{r_{bot}(t)} c_g(r) \cdot 2\pi r \cdot dr,
\end{align}
where $r_{top}(t)$ and $r_{bot}(t)$ are the radii for the top and bottom apertures that the fuel can escape through, and $c_g(r)$ is the hydrodynamic sound speed, given by
\begin{equation}
c_g(r) = \sqrt{\gamma_g p_g/\rho_g(r)},
\end{equation}
where $\gamma_g=5/3$ is the ratio of specific heats for an ideal gas.  The factor of $(3/4)^4$ is a result of the derivation presented in Ref.~\onlinecite{Slutz_PoP_2010}.  Finally, the total energy loss due to end losses is\cite{endlossnote}
\begin{equation}
\dot{E}_{ends} = \dot{E}_{top} + \dot{E}_{bot}.
\end{equation}

Similarly, for mass losses from the fuel, we assume ion density flows across the apertures, and thus, for ion species $s$, we have
\begin{align}
\dot{N}_{s,top} &= (3/4)^4  \int_0^{r_{top}(t)} n_s(r)  c_g(r) \cdot 2\pi r \cdot dr \\
\dot{N}_{s,bot} &= (3/4)^4  \int_0^{r_{bot}(t)} n_s(r)  c_g(r) \cdot 2\pi r \cdot dr \\
\dot{N}_{s,ends} &= \dot{N}_{s,top} + \dot{N}_{s,bot}.
\end{align}

The number of deuterons and tritons in the fuel change because of fusion reactions as well as end losses (see Eqs.~\ref{Nddot} and \ref{Ntdot} below).  By contrast, end losses are the only way for prescribed dopant and/or contaminant particle numbers to change.  Thus, we have
\begin{equation}\label{Nmixdot}
\dot{N}_{mix} = -\sum_{s\neq d,t}\dot{N}_{s,ends},
\end{equation} 
where $N_{mix}(t) = \sum_{s\neq d,t} N_s(t)$ is the total number of dopant and contaminant particles.  Note that we only need one differential equation to describe the evolution of all of the ``mix" particles because their number distribution amongst themselves remains constant in time, i.e.,
\begin{equation}
f_{s\neq d,t}(t) \equiv \frac{N_{s}(t)}{N_{mix}(t)} = const. = f_{s}(t_0),
\end{equation}
and therefore
\begin{equation}
N_{s\neq d,t}(t) = f_{s}(t_0) \cdot N_{mix}(t).
\end{equation}

Finally, should fuel preheating be done with a laser entering from above the imploding region, then these end loss calculations should be made for at least the top plane, where the radius of the top aperture is given by
\begin{equation}
r_{top}(t) = \min{\left\{r_{LEC}, r_g(t) \right\}},
\end{equation}
and where $r_{LEC}\geq r_b$ is the radius of the laser entrance channel.  The calculation for the bottom plane, however, depends on whether the bottom plane is completely closed off with electrode material (i.e., an implosion ``glide plane") or if it has an opening for a beam dump.  Should a beam dump be used, then the radius of the bottom aperture is given by
\begin{equation}
r_{bot}(t) = \min{\left\{r_{dump}, r_g(t) \right\}},
\end{equation}
where $r_{dump}$ is the radius of the beam dump.

\subsection{\label{sec:circuit}Magnetic flux loss due to the Nernst thermoelectric effect}

With all of the terms in Eq.~\ref{Egdot} thus defined, the next important quantity to calculate is the flux loss of the axial magnetic field from the fuel region to the liner region due to the Nernst thermoelectric effect. For MagLIF with a hot fuel region, flux losses due to the Nernst effect dominate over flux losses due to resistive diffusion alone.\cite{Slutz_PoP_2010} To model flux losses due to the Nernst effect, we follow Ref.~\onlinecite{Braginskii_1965} to write
\begin{align}\label{Phizgdot}
\dot{\Phi}_{zg} &= \left[-2\pi r\cdot \mathcal{F}\left(x_e\right)\cdot \frac{k}{q_e}\frac{\partial T_{g}}{\partial r}\right]_{r=r_g} \\
\mathcal{F}(x_e) &= \frac{1.5x_e^3+3.053x_e}{x_e^4+14.79x_e^2+3.7703}.
\end{align}
Note that, like thermal conduction, Nernst losses also depend on a function of the Hall parameter, $\mathcal{F}(x_e\equiv \omega_{ce}\tau_{ei})$, as well as gradients in the fuel temperature. Also, we assume that the magnetic flux lost by the fuel is transported to the liner, and thus
\begin{equation}
\Phi_{zl}(t) = \Phi_{zl0} + \left[\Phi_{zg0} - \Phi_{zg}(t) \right].
\end{equation}
As the axial flux is transported into the liner, it is likely that it is dissipated rapidly by resistive diffusion, which would heat the liner material somewhat.\cite{Velikovich_personal_2014}  However, the energy density in the liner for either case (either pure magnetic energy or ohmic dissipation leading to thermal energy) is roughly the same, and thus will affect the liner dynamics similarly given the other simplifying assumptions made in our model.  Thus, for simplicity, we ignore ohmic dissipation of the axial magnetic field.  Finally, we assume that the axial magnetic flux in the vacuum region is conserved, thus $\Phi_{zv}(t)=\Phi_{zv0}$.

\subsection{\label{sec:erosion}Erosion of the fuel's dense outer shelf region}

The cold dense shelf region, generated by preheating only a central portion of the fuel, provides a buffer region in the fuel between the hot spot and the cold liner wall.  This buffer region significantly reduces radiation losses by effectively reducing the amount of fuel mass that contributes to the radiation losses.  It also reduces thermal conduction losses and axial magnetic flux losses from the fuel to the liner because the temperature gradient at the edge of the nearly flat shelf region is much less than that at the edge of the hot spot region.  However, this buffer region is not static.  It begins to erode away immediately after its formation due to thermal conduction from the hot spot region to the shelf region.  To describe this erosion, we differentiate $E=\frac{3}{2}NkT$ with respect to time, rearrange the terms, and make simple substitutions to approximate the mass transfer rate from the shelf region to the hot spot region as
\begin{equation}\label{mdot}
\dot{m}_{s\to h}=\frac{2}{3}\frac{\bar{m}_i \left(P_{ch} - P_{cs}\right)}{\left(1+\bar{Z}_g\right) k(\bar{T}_{h}-\bar{T}_{s})},
\end{equation}
where $\bar{T}_h$ is the average temperature in the hot spot region and $\bar{T}_s$ is the average temperature in the shelf region. This states that thermal energy must be deposited in the shelf for a particle to make a ``quantum" jump from having an average thermal energy of $\frac{3}{2}k\bar{T}_s$ in the shelf region to an average thermal energy of $\frac{3}{2}k\bar{T}_h$ in the hot spot region; the larger the temperature jump and/or mass transfer rate, the larger $P_{ch}-P_{cs}$ needs to be. The average temperatures in each region are found by invoking our isobaric assumption, i.e.,
\begin{align}
\bar{T}_h &= \bar{\rho}_g \bar{T}_g / \bar{\rho}_h \\
\bar{T}_s &= \bar{\rho}_g \bar{T}_g / \bar{\rho}_s,
\end{align}
where
\begin{align}
\bar{\rho}_h &= f_{mh} m_g / \left(\pi r_h^2 h\right)  \\
\bar{\rho}_s &= \left(1-f_{mh}\right) m_g / \left[\pi \left(r_g^2 - r_h^2\right) h\right],
\end{align}
are the mean fuel densities in the hot spot and shelf regions, respectively, and $f_{mh}$ is the fraction of the total fuel mass that is part of the hot spot.  We keep track of $f_{mh}$ rather than the absolute fuel mass in the hot spot since the absolute mass changes (due to fusion reactions and end losses). Thus we have
\begin{equation}\label{fmhdot}
\dot{f}_{mh}=\dot{m}_{s\to h}/m_g.
\end{equation}
Note that initially (at the time of preheat), we have
\begin{equation}\label{fmh0}
f_{mh}(t_{ph})=\left[ \frac{r_{ph0}}{r_g(t_{ph})}  \right]^2,
\end{equation}
where if laser absorption is being calculated, $r_{ph0} = r_b$.  From our example above using the laser preheating parameters of recent MagLIF experiments,\cite{Gomez_PRL_2014} we get
\begin{equation}
f_{mh}(t_{ph})=\left[ \frac{0.492}{2.325}  \right]^2 = \left[ 0.21 \right]^2 = 0.04.
\end{equation}
Thus, initially, only about 4\% of the fuel mass is part of the hot spot region, though, radially, the hot spot region occupies 21\% of the overall fuel radius $r_g$, and quickly expands to occupy $\sim$ 90\% of $r_g$ before the shelf begins eroding away with any significance.

\subsection{\label{sec:circuit}Primary fusion reaction rates}

For an arbitrary deuterium to tritium fuel ratio, the DT reaction rate is given by\cite{Harms_2000}
\begin{equation}\label{Ndtdot}
\dot{N}_{dt}=h\int_0^{r_g}{n_d n_t \langle\sigma v\rangle_{dt} \cdot 2\pi r \cdot dr},
\end{equation}
and the two dominant DD reaction rates are given by\cite{Harms_2000} 
\begin{align}
\label{Ndd3Hedot} \dot{N}_{dd,^3{\rm He}}&=\frac{h}{2}\int_0^{r_g}{n_d^2\langle\sigma v\rangle_{dd,^3{\rm He}}  \cdot 2\pi r \cdot dr} \\
\label{Nddtdot} \dot{N}_{dd,t}&=\frac{h}{2}\int_0^{r_g}{n_d^2\langle\sigma v\rangle_{dd,t}  \cdot 2\pi r \cdot dr},
\end{align}
where $n_d$ and $n_t$ are the temporally and radially dependent deuteron and triton number densities in the fuel, as defined by Eq.~\ref{eqn:n_s} with $s=d$ and $s=t$, respectively, and where $\langle\sigma v\rangle_{dt}$, $\langle\sigma v\rangle_{dd,{\rm ^3He}}$, and $\langle\sigma v\rangle_{dd,t}$ are the temporally and radially dependent reactivity parameters for the DT, DD-$^3$He, and DD-T reactions, respectively. The reactivity parameters are calculated using\cite{Bosch_and_Hale_NF_1992,Atzeni_2004}
\begin{subequations}\label{sigv}
\begin{align}
\langle\sigma v\rangle &= C_1\zeta^{-5/6}\xi^2\exp{(-3\zeta^{1/3}\xi)}\\
\zeta&=1-\frac{C_2T_{g,keV}+C_4T_{g,keV}^2+C_6T_{g,keV}^3}{1+C_3T_{g,keV}+C_5T_{g,keV}^2+C_7T_{g,keV}^3}\\
\xi&=C_0/T_{g,keV}^{1/3},
\end{align}
\end{subequations}
where the coefficients $C_0$--$C_7$ are the fitting parameters provided in Table~\ref{sigvcoefs}.
\begin{table}
\caption{\label{sigvcoefs}Coefficients for Eqs.~\ref{sigv} (from Refs.~\onlinecite{Bosch_and_Hale_NF_1992,Atzeni_2004}).}
\begin{ruledtabular}
\begin{tabular}{lccc}
      & DT & DD,$^3$He & DD,T \\
\hline
$C_0$ & 6.6610 & 6.2696 & 6.2696 \\
$C_1\times 10^{22}$ & $643.41\times 0.98$\footnotemark[1] & 3.5741\footnotemark[2] & 3.7212 \\
$C_2\times 10^{3}$ & 15.136 & 5.8577 & 3.4127 \\
$C_3\times 10^{3}$ & 75.189 & 7.6822 & 1.9917 \\
$C_4\times 10^{3}$ & 4.6064 & 0 & 0 \\
$C_5\times 10^{3}$ & 13.500 & -0.002964\footnotemark[2] & 0.010506 \\
$C_6\times 10^{3}$ & -0.10675 & 0 & 0 \\
$C_7\times 10^{3}$ & 0.01366 & 0 & 0 \\
\end{tabular}
\end{ruledtabular}
\footnotetext[1]{For DT, we have multiplied the $C_1$ coefficient of Ref.~\onlinecite{Atzeni_2004} by 0.98 for a slightly better fit to the peak of the tabulated $\langle\sigma v\rangle$ data published in Ref.~\onlinecite{McNally_ORNL_1979}.}
\footnotetext[2]{For DD,$^3$He, and for $T>100$ keV, we set $C_5=0$ to avoid $\langle\sigma v\rangle_{dd,^3{\rm He}}\to\pm\infty$ near 965 keV.  Though unlikely for any reasonable MagLIF case, this $\pm\infty$ has caused computational problems when scanning over a large parameter space, where very low density plasmas can be heated to 1 MeV.  This replacement causes a discontinuity at 100 keV; to remove this discontinuity, we also set $C_1\times 10^{22} = 3.5741 \times 1.0172$ for $T>100$ keV.}
\end{table}

The reactivity rates, along with end losses, provide the rate of change of the total number of deuterons and tritons in the fuel, respectively, as
\begin{align}
\dot{N}_d &= -\dot{N}_{dt}-2\dot{N}_{dd,^3He}-2\dot{N}_{dd,t}-\dot{N}_{d,ends} \label{Nddot} \\
\dot{N}_t &= -\dot{N}_{dt}-\dot{N}_{t,ends}. \label{Ntdot}
\end{align}
The reactivity rates also provide the total fusion power, as given by
\begin{equation}\label{Pdt}
P_{f}=\dot{N}_{dt} Q_{dt}+\dot{N}_{dd,{\rm ^3He}} Q_{dd,{\rm ^3He}}+\dot{N}_{dd,t} Q_{dd,t},
\end{equation}
where the energy yields per DT, DD-$^3$He, and DD-T reaction are, respectively,
\begin{subequations}
\begin{align}
Q_{dt}&=17.6\times 10^6\cdot q_e \label{Qdt}\\
Q_{dd,{\rm ^3He}}&=3.27\times 10^6\cdot q_e \label{Qdd3He}\\
Q_{dd,t}&=4.03\times 10^6\cdot q_e. \label{Qddt}
\end{align}
\end{subequations}
Since one neutron is released per DT reaction and one neutron is released per DD-$^3$He reaction, the primary DT and primary DD neutron yields are given by
\begin{align}
\label{YDTn} Y_{dt,n}&=N_{dt} \\
\label{YDDn} Y_{dd,n}&=N_{dd,{\rm ^3He}},
\end{align}
and thus the total primary neutron yield is
\begin{equation}
\label{Yn} Y_{n}=Y_{dt,n}+Y_{dd,n}.
\end{equation}
Finally, the total fusion energy yield is
\begin{equation}\label{Pdt}
Y={N}_{dt} Q_{dt}+{N}_{dd,{\rm ^3He}} Q_{dd,{\rm ^3He}}+{N}_{dd,t} Q_{dd,t}.
\end{equation}

\subsection{\label{sec:circuit}Summary of the semi-analytic MagLIF model}

{\allowdisplaybreaks
The dynamics of our model are described by Eqs.~\ref{Isdot}, \ref{Vcdot}, \ref{Ildot}, \ref{rliddot}, \ref{rgddot}, \ref{rlddot}, \ref{Elsdot}, \ref{Egdot}, \ref{Nmixdot}, \ref{Phizgdot}, \ref{fmhdot}, \ref{Ndtdot}, \ref{Ndd3Hedot}, \ref{Nddtdot}, \ref{Nddot}, and \ref{Ntdot}.  These $2N_{ls}+13$ ordinary differential equations are repeated here for emphasis and clarity:
\begin{subequations}
\begin{align}
&\dot{I}_s = \frac{\varphi_{oc} - Z_0I_{s} - \varphi_c}{L} \label{Isdot2} \\
&\dot{\varphi}_c = \frac{I_s-I_l-\varphi_c/R_{loss}}{C} \\
&\dot{I}_l = \frac{\varphi_c-I_l(\dot{L}_v + \dot{L}_{lc})}{L_0+L_v+L_{lc}} \label{Ildot2} \\
&\ddot{r}_{l,i} = \frac{p_{l,s=i-1}-p_{l,s=i}}{m_{ls}} \cdot 2\pi r_{l,i} \cdot h \\ 
		&\phantom{=} \quad\quad\quad\quad\quad\quad\quad\quad\quad\quad\quad \left(i=2,3,\ldots,N_{li}-1\right) \nonumber \\
&\ddot{r}_g = \frac{p_{g}+p_{\bar{B}_{zg}}-p_{l,s=1}}{m_{ls}/2}\cdot 2\pi r_g \cdot h \\
&\ddot{r}_l = \frac{p_{l,s=N_{ls}}-p_{B_{\theta lv}}-p_{\bar{B}_{zv}}}{m_{ls}/2}\cdot 2\pi r_l \cdot h \\
&\dot{E}_{l,s} = -\left( \frac{2}{3}\frac{E_{l,s}}{V_{l,s}} + q_{l,s} \right)\dot{V}_{l,s} \\
		&\phantom{=} \quad\quad + \left( P_r + P_c + P_{\Omega} - P_{BB} \right)/N_{ls} \nonumber \\ 
		&\phantom{=} \quad\quad\quad\quad\quad\quad\quad\quad\quad\quad\quad\quad \left(s=1,2,\ldots,N_{ls}\right) \nonumber \\
&\dot{E}_g = P_{pdV} + P_{ph} + P_{\alpha} - P_r - P_c - \dot{E}_{ends} \\
&\dot{N}_{mix} = -\sum_{s\neq d,t}\dot{N}_{s,ends} \\
&\dot{\Phi}_{zg} = \left[-2\pi r\cdot \mathcal{F}\left(x_e\right)\cdot \frac{k}{q_e}\frac{\partial T_{g}}{\partial r}\right]_{r=r_g} \\
&\dot{f}_{mh} = \dot{m}_{s\to h}/m_g \label{fmhdot2}\\
&\dot{N}_{dt} = h\int_0^{r_g}{n_d n_t \langle\sigma v\rangle_{dt} \cdot 2\pi r \cdot dr} \\
&\dot{N}_{dd,^3{\rm He}} = \frac{h}{2}\int_0^{r_g}{n_d^2\langle\sigma v\rangle_{dd,^3{\rm He}}  \cdot 2\pi r \cdot dr} \\
&\dot{N}_{dd,t} = \frac{h}{2}\int_0^{r_g}{n_d^2\langle\sigma v\rangle_{dd,t}  \cdot 2\pi r \cdot dr} \\
&\dot{N}_d = -\dot{N}_{dt}-2\dot{N}_{dd,^3He}-2\dot{N}_{dd,t}-\dot{N}_{d,ends}\\
&\dot{N}_t = -\dot{N}_{dt}-\dot{N}_{t,ends}.
\end{align}
\end{subequations}}

Note that if the liner current is prescribed, rather than derived from the voltage-driven circuit model, then Eqs.~\ref{Isdot2}--\ref{Ildot2} are discarded, and only $2N_{ls}+10$ ordinary differential equations are required to describe the dynamics of our model. Furthermore, if we ignore the Nernst effect, preheat all of the fuel uniformly, use either pure D$_2$ fuel or a 50/50\% DT mix, and we ignore dopants and contaminants, then the number of required differential equations drops to only $2N_{ls}+5$.

\section{\label{sec:results}Model verification}

\begin{figure*}
\includegraphics[width=1.0\textwidth]{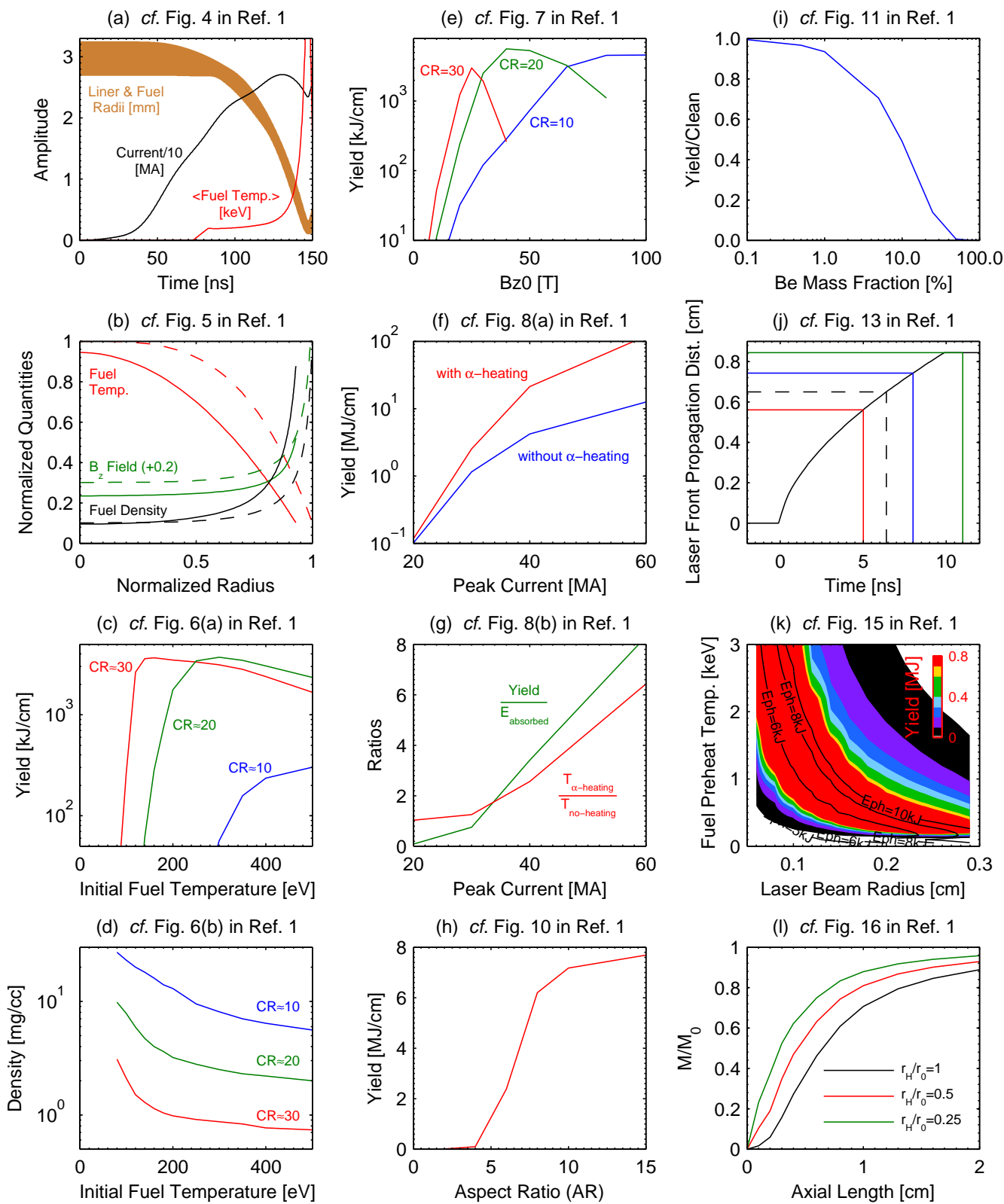}
\caption{\label{fig_code_verification} Simulation results from our semi-analytic MagLIF model (SAMM), where the intent was to reproduce the simulation results presented in the corresponding figures of Ref.~\onlinecite{Slutz_PoP_2010}.}
\end{figure*}

In Fig.~\ref{fig_code_verification}, we present SAMM simulation results where the intent was to reproduce the 1D results presented in the original MagLIF paper.\cite{Slutz_PoP_2010} These SAMM results show that our simplified model does indeed capture the 1D behavior presented throughout Ref.~\onlinecite{Slutz_PoP_2010}.  All of the simulations presented in Fig.~\ref{fig_code_verification} used DT fuel.

In Fig.~\ref{fig_code_verification}(a), we have plotted the fuel radius, liner radius, drive current, and average fuel temperature as a function of time for the preliminary point design discussed in Ref.~\onlinecite{Slutz_PoP_2010}. This figure should be compared with Fig.~4 in Ref.~\onlinecite{Slutz_PoP_2010}.  The SAMM simulation produced a fusion energy yield of 970 kJ, compared to about 500 kJ in Ref.~\onlinecite{Slutz_PoP_2010}, and resulted in a maximum convergence ratio of 25, which matches the preliminary point design of Ref.~\onlinecite{Slutz_PoP_2010}.

In Fig.~\ref{fig_code_verification}(b), we have plotted the normalized fuel temperature, magnetic field strength, and fuel density at stagnation (at peak burn rate) as a function of the normalized fuel radius for two SAMM simulations, one with the Nernst effect included (solid lines), and one without the Nernst effect included (dashed lines). All quantities are normalized to the simulation without the Nernst effect.\cite{Bzshift}  This figure should be compared with Fig.~5 in Ref.~\onlinecite{Slutz_PoP_2010}.  The yield for the simulation with Nernst is 72\% lower than the simulation without Nernst (compared to about 70\% in Ref.~\onlinecite{Slutz_PoP_2010}).  The simulation with Nernst resulted in an axial magnetic flux loss of about 64\% (compared to about 70\% in Ref.~\onlinecite{Slutz_PoP_2010}).

Figures~\ref{fig_code_verification}(c) and \ref{fig_code_verification}(d) should be compared with Figs.~6(a) and 6(b) in Ref.~\onlinecite{Slutz_PoP_2010}.  In Ref.~\onlinecite{Slutz_PoP_2010}, the initial fuel densities that result in solutions with convergence ratios of 10, 20, and 30, were found as a function of the initial fuel temperature.  Using these initial conditions directly with SAMM, shown here in Fig.~\ref{fig_code_verification}(d), we generated the results shown in Fig.~\ref{fig_code_verification}(c).  The resulting convergence ratios were 16.3$\pm$1.4, 25.7$\pm$3.4, and 32$\pm$5.5, and thus are only approximately equal to 10, 20, and 30.

In Fig.~\ref{fig_code_verification}(e), we have plotted the optimum fusion energy yields per unit liner length as a function of the initial axial magnetic field strength.  Here we have held the maximum convergence ratio fixed at either 10, 20, or 30.  To find these optimum constant convergence ratio solutions, we had to scan over the input parameter space that consists of the initial preheat energy and the initial fuel density.  These results should be compared with Fig.~7 in Ref.~\onlinecite{Slutz_PoP_2010}.

Figures~\ref{fig_code_verification}(f) and \ref{fig_code_verification}(g) should be compared with Figs.~8(a) and 8(b) in Ref.~\onlinecite{Slutz_PoP_2010}.  In Fig.~\ref{fig_code_verification}(f), we have plotted the optimum fusion energy yields per unit liner length as a function of the peak drive current for two cases, one with $\alpha$-deposition turned on, and one with $\alpha$-deposition turned off.  Here, as in Ref.~\onlinecite{Slutz_PoP_2010}, for the case with $\alpha$-heating, we scanned over the input parameter space that consists of the initial preheat energy and the initial fuel density to find the optimum solutions that had a maximum convergence ratio of 20 (for the case with no $\alpha$-heating, the same input parameters were used, but $\alpha$-deposition was simply turned off, thus the convergence ratios varied somewhat).  In Fig.~\ref{fig_code_verification}(g), we have plotted the ratio of the peak fuel temperature with $\alpha$-heating turned on to the peak fuel temperature with $\alpha$-heating turned off.  We have also plotted the ratio of the total fusion energy yield to the total energy absorbed by the target (liner and fuel), i.e., the ``target gain''.

In Fig.~\ref{fig_code_verification}(h), we have plotted the optimum fusion energy yields per unit liner length as a function of the initial liner aspect ratio (AR).  This figure should be compared with Fig.~10 in Ref.~\onlinecite{Slutz_PoP_2010}.  In the SAMM results, the transition to higher yields with higher aspect ratios is a bit more abrupt, and then rolls off sooner; however, the general trend is the same as that shown in Ref.~\onlinecite{Slutz_PoP_2010}.  Also, although not explicitly shown in Ref.~\onlinecite{Slutz_PoP_2010}, the eventual rolloff in yield with higher aspect ratio is expected.  That is, to keep the implosion time constant with higher aspect ratio, the overall liner mass must be decreased while the liner's initial outer radius is increased (and while the liner's initial wall thickness is decreased).  Eventually, the liner's mass decreases to a point where it cannot provide sufficient tamping and inertial confinement at stagnation.

In Fig.~\ref{fig_code_verification}(i), we have plotted the normalized yield as a function of the premixed fraction of beryllium in the fuel.  The yields are normalized to the case with no mix/dopants.  This figure should be compared with Fig.~11 in Ref.~\onlinecite{Slutz_PoP_2010}.

In Fig.~\ref{fig_code_verification}(j), we consider laser propagation through the fuel using the analytic treatment discussed in Sec.~\ref{sec:preheat}.  Here we plot the laser heating front propagation distance as a function of time for the preliminary point design of Ref.~\onlinecite{Slutz_PoP_2010}.  The laser pulse duration is 10 ns.  At 5 ns (red), the front of the ``bleaching wave'' has reached 0.56 cm into the fuel, which is consistent with Fig.~13(b) of Ref.~\onlinecite{Slutz_PoP_2010}.  By 8 ns (blue) and 11 ns (green), the bleaching wave has propagated completely through the 0.65-cm-long imploding fuel region, which is consistent with Figs.~13(c) and 13(d) of Ref.~\onlinecite{Slutz_PoP_2010}, respectively.

In Fig.~\ref{fig_code_verification}(k), we present a color-filled contour plot of total fusion energy yield as a function of both the fuel preheat temperature and the laser beam spot size (radius).  Overlaid on top of the color-filled contour plot are the contours of constant preheat energy required to obtain the preheat temperature at the given spot sizes.  This figure should be compared with Fig.~15 in Ref.~\onlinecite{Slutz_PoP_2010}.  In Ref.~\onlinecite{Slutz_PoP_2010}, this figure was generated by additionally scanning over a range of initial fuel densities to obtain only the solutions where the convergence ratios were 25.  For simplicity, we used the input densities and preheat energies found by this previous effort to generate the SAMM results presented here in Fig.~\ref{fig_code_verification}(k), and thus the convergence ratios were allowed to vary somewhat.

Finally, in Fig.~\ref{fig_code_verification}(l), we have plotted results for mass end losses out of the fuel region through the laser entrance hole using the analytic treatment discussed in Sec.~\ref{sec:endlosses}.  Here we are plotting the ratio of mass remaining after fusion burn to the mass of the fuel at the start of the simulation as a function of the axial length of the liner.  We have done this for three cases, where the ratio of the radius of the laser entrance hole to the initial radius of the fuel is either 1.0, 0.5, or 0.25.  This figure should be compared with Fig.~16 in Ref.~\onlinecite{Slutz_PoP_2010}.

\section{\label{sec:summary}Summary, Future Work, and Conclusions}

In this article, we have presented a new semi-analytic model of MagLIF.  This model, implemented in a code called SAMM, accounts for: (1) preheat of the fuel (optionally via laser absorption); (2) pulsed-power-driven liner implosion; (3) liner compressibility with an analytic equation of state, artificial viscosity, internal magnetic pressure, and ohmic heating; (4) adiabatic compression and heating of the fuel; (5) radiative losses and fuel opacity; (6) magnetic flux compression with Nernst thermoelectric losses; (7) magnetized electron and ion thermal conduction losses; (8) end losses; (9) enhanced losses due to prescribed dopant concentrations and contaminant mix; (10) deuterium-deuterium and deuterium-tritium primary fusion reactions for arbitrary deuterium to tritium fuel ratios; and (11) magnetized $\alpha$-particle heating.  We have also shown that SAMM is capable of reproducing the general 1D behavior presented throughout the original MagLIF paper (i.e., Ref.~\onlinecite{Slutz_PoP_2010}).

Presently, we are using SAMM to further explore the parameter space of MagLIF.  These efforts are complementary to other studies presently taking place using full 1D, 2D, and 3D radiation magnetohydrodynamics codes.\cite{Jennings_personal_2015,Sefkow_personal_2015,Slutz_personal_2015}  In particular, we are using SAMM to study the parameter space surrounding the experimental platform of Ref.~\onlinecite{Gomez_PRL_2014} (i.e., an initial axial magnetic field of 10 T, a preheat energy of less than 2 kJ, an initial fuel density of about 1 mg/cm$^3$, and a peak drive current of about 20 MA).  We are also studying how this parameter space could change as a series of planned upgrades are made to the Z facility over the next few years.  These upgrades are intended to bring MagLIF's experimental platform closer to the parameters of the preliminary point design described in Ref.~\onlinecite{Slutz_PoP_2010} (i.e., an initial axial magnetic field of 30 T, a preheat energy of 8 kJ, an initial fuel density of 3 mg/cm$^3$, and a peak drive current of 27 MA).  Additionally, we are studying how this parameter space could change as the peak drive current is increased to about 60 MA, in consideration of MagLIF on possible future pulsed-power accelerators, such as the conceptual Z-300 and Z-800 accelerators.\cite{Stygar_personal_2015}  The results of these parameter space studies using SAMM will be presented in a future publication.

In closing, we note that the development of SAMM has led to several physical insights that were not fully appreciated previously (e.g., the dependence of radiative loss rates on the radial fraction of the fuel that is preheated).  Additionally, this model's accessible physics and fast run times ($\sim$~30 seconds/simulation) make it a useful pedagogical tool, especially for students, experimentalists, or any researcher interested in MagLIF.

\begin{acknowledgments}
We would like to thank D.~B.~Sinars for supporting this project and reviewing this manuscript.  We also thank M.~E.~Cuneo, M.~C.~Herrmann, C.~W.~Nakhleh, K.~J.~Peterson, and M.~K.~Matzen for supporting this project.  We thank R.~A.~Vesey for useful discussions about $\langle\sigma v\rangle$ approximations, MagLIF parameter scans, and for supplying the simulation input parameters needed for Fig.~\ref{fig_code_verification}(k) in this manuscript.  We thank K.~R.~Cochrane for supplying us with SESAME equation-of-state data and for suggesting the use of the Birch-Murnaghan formulation as we attempted to fit the SESAME cold-curve data to various analytic formulae.  We thank S.~B.~Hansen for useful discussions about various simplified radiation loss models.  We thank P.~F.~Schmit, P.~F.~Knapp, and M.~R.~Gomez for some helpful beta testing of the SAMM software.  We thank M.~R.~Gomez for useful discussions about fully-integrated MagLIF experiments.  We thank M. Geissel for useful discussions about laser transmission through laser entrance windows.  We thank the MagLIF/ICF and Dynamic Material Properties research groups in general for many useful technical discussions regarding various aspects of MagLIF; in particular, we thank A.~B.~Sefkow, M.~R.~Martin, C.~A.~Jennings, R.~W.~Lemke, T.~J.~Awe, D.~C.~Rovang, D.~C.~Lamppa, and W.~A.~Stygar.  Finally, we thank the personnel of the Pulsed-Power Sciences Center (including the Z and ZBL facilities) at Sandia National Laboratories; without their hard work and dedication to excellence, this work would not have been possible.

Sandia is a multiprogram laboratory operated by Sandia Corporation, a Lockheed-Martin company, for the United States Department of Energy's National Nuclear Security Administration, under Contract No. DE-AC04-94AL85000.
\end{acknowledgments}

\appendix

\section{\label{loglambda}Calculations of the Coulomb logarithm, $\ln\Lambda$}

Following Ref.~\onlinecite{Callen}, the Coulomb logarithm is given by
\begin{equation}
\ln\Lambda = \max\left\{1,\ln\left(\frac{\lambda_{D}}{b_{min}}\right)\right\},
\end{equation}
where $\lambda_D$ is the Debye length of the plasma and $b_{min}$ is the minimum impact parameter of the plasma.  The Debye length is given by
\begin{equation}
\lambda_D = \sqrt{\frac{\epsilon_0 kT}{q_e^2\left(n_e+\sum_s{\bar{Z}_s^2 n_s}\right)}}.
\end{equation}
The minimum impact parameter is conventionally taken as
\begin{equation}
b_{min} = \max \left\{ b_{min}^{cl}, b_{min}^{qm} \right\},
\end{equation}
where
\begin{equation}
b_{min}^{cl} = \frac{\bar{Z}_g q_e^2}{\left\{ 4\pi\epsilon_0 \right\} 3 kT}
\end{equation}
is the classical minimum impact parameter, and
\begin{equation}
b_{min}^{qm} = \frac{\hbar}{2 m_e v_{T_e}}
\end{equation}
is the quantum mechanical minimum impact parameter.  Here, $\hbar = 1.05457 \times 10^{-34}$ $\rm J\cdot s$ is the reduced Planck constant and
\begin{equation}
v_{T_e} = \sqrt{2 k{T}/m_e}
\end{equation}
is the electron thermal velocity.

For the $\ln\Lambda$ calculation in Eq.~\ref{nueizblabs}, we use the following assignments: $\bar{Z}_s\to Z_{nuc,s}$, $kT\to13.6\cdot q_e \cdot \sum_s{Z_{nuc,s}^\zeta}$, $n_e\to \bar{n}_e$, and $n_s\to \bar{n}_s$.

For the $\ln\Lambda$ calculation in Eq.~\ref{lalpha}, we use the following assignments: $kT\to k\bar{T}_g$, $n_e\to \bar{n}_e$, and $n_s\to \bar{n}_s$.

For the $\ln\Lambda$ calculations in Eqs.~\ref{nueic} and \ref{nuiic}, we use the following assignments: $kT\to kT_g(r)$, $n_e\to n_e(r)$, and $n_s\to n_s(r)$.

\section{\label{rhnotes}Additional computational details for finding $r_h$, $T_B$, $\rho_c$, and $T_c$}
\subsection{When the shelf region is \underline{not} present}
This case occurs when either all of the fuel is preheated uniformly, or when the shelf region has completed eroded away (see Sec.~\ref{sec:erosion}).  For this case, $r_h=r_g$, and we iteratively guess for $T_B$ in our bisection algorithm.  After evaluating the expressions in Eqs.~\ref{Prh1}--\ref{Prh2}, if $P_{rs}(r_h)<P_{rv}(r_h)$, then $T_B$ is too low, and the correct value for $T_B$ will be found between this value that is too low and the average temperature of the entire fuel, $\bar{T}_g$.  [Note that we take $\bar{T}_g$ as the initial upper limit for $T_B$ in the bisection algorithm because with $r_h=r_g$ and $T_B=\bar{T}_g$, the resulting fuel temperature and density profiles are flat, and $P_{rv}(r_h)=0$ while $P_{rs}(r_h)>0$.]  Conversely, if $P_{rs}(r_h)>P_{rv}(r_h)$, then $T_B$ is too high, and the correct value for $T_B$ will be found between this value that is too high and zero. [Note that $T_B=0$ results in $P_{rs}(r_h)=0$ while $P_{rv}(r_h)>0$, thus we take zero as the initial lower limit for $T_B$ in the bisection algorithm.]

\subsection{When the shelf region is present}
For this case, we must ensure that $r_h$ and $T_B$ are chosen such the following mass conservation argument is met:
\begin{align}
m_{gs}=(1-f_{mh})m_g &= \int_{r_h}^{r_g} \rho_g(r) \cdot 2\pi r \cdot h \cdot dr \\
				&= \frac{4}{9}\cdot \frac{2\pi h \rho_c T_c}{T_B r_h^{\nicefrac{1}{4}}} \cdot \left( r_g^{\nicefrac{9}{4}} - r_h^{\nicefrac{9}{4}} \right) \\
				&= \frac{4}{9}\cdot \frac{2\pi h \bar{\rho}_g \bar{T}_g}{T_B r_h^{\nicefrac{1}{4}}} \cdot \left( r_g^{\nicefrac{9}{4}} - r_h^{\nicefrac{9}{4}} \right). \label{B3}
\end{align}
Here $f_{mh}$ is the fraction of the total fuel mass that is in the hot spot region, which is the system variable defined by Eqs.~\ref{fmhdot}--\ref{fmh0} (see also Eq.~\ref{fmhdot2}). Also, we have used Eq.~\ref{rhogrshelf} for $\rho_g(r)$ and, in the last equality, we have used our isobaric assumption that $\rho_c T_c = \bar{\rho}_g \bar{T}_g$.  Equation~\ref{B3} provides the analytic relationship between $T_B$ and $r_h$ that ensures mass conservation in the shelf region.  However, if we were to iteratively guess for $T_B$ in our bisection algorithm, then solving Eq.~\ref{B3} for $r_h$ in terms of $T_B$ would require a numerical treatment.  Therefore, we instead guess iteratively for $r_h$, which must reside in the domain $0<r_h<r_g$ (thus, for our bisection algorithm, we use $r=0$ and $r=r_g$ as our initial lower and upper limits, respectively).  For each $r_h$ guess, Eq.~\ref{B3} provides the unique value of $T_B$ that will ensure mass conservation.  Note that $T_B(r_h)$ is a monotonically decreasing function (i.e., $T_B\to \infty$ as $r_h\to 0$, and $T_B\to 0$ as $r_h\to r_g$).  Thus, in our bisection algorithm, if $P_{rs}(r_h)<P_{rv}(r_h)$, then $T_B$ is too low and $r_h$ is too large; the correct value for $r_h$ will be found between this value that is too large and zero.  Conversely, if $P_{rs}(r_h)>P_{rv}(r_h)$, then $T_B$ is too high and $r_h$ is too small; the correct value for $r_h$ will be found between this value that is too small and $r_g$.

\subsection{Updating $\rho_c$ and $T_c$}
Before we can test our iterative guesses for $r_h$ and $T_B$ in Eqs.~\ref{Prh1}--\ref{Prh2}, we first need to find $\rho_c$ and $T_c$ to complete the profile definitions given in Eqs.~\ref{Tgr}--\ref{rhogrshelf}.  To do this, we use our isobaric assumption to first update $\rho_g(r_h)=\bar{\rho}_g\bar{T}_g/T_B\equiv \rho_B$.  Next, we find the value for $\rho_c$ that makes the total mass in the hot spot, as defined by Eq.~\ref{rhogr}, consistent with $m_{gh}=f_{mh}m_g$ (see Appendix~\ref{rhocnotes} for additional details).  Then, we again apply our isobaric assumption to get $T_c=\bar{\rho}_g\bar{T}_g/\rho_c$.  Finally, with all profile parameters defined and updated, we evaluate $P_{rv}(r_h)$ and $P_{rs}(r_h)$ and test whether the results meet our desired solution accuracy.

\subsection{\label{rhocnotes}Additional details for updating $\rho_c$}
\begin{figure}
\includegraphics[width=\columnwidth]{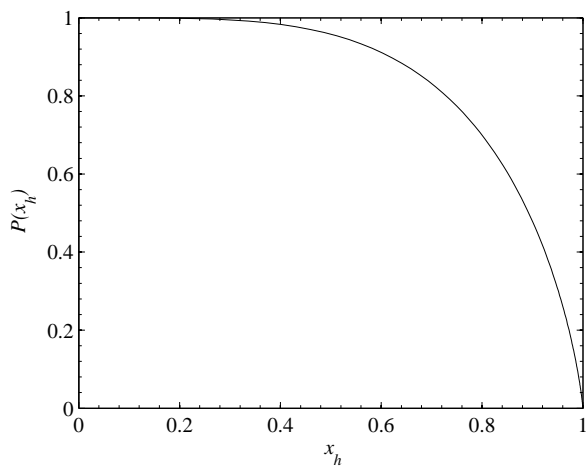}
\caption{\label{Pxh}Plot of $\mathcal{P}(x_h)$, the dimensionless, monotonically decreasing function of the dimensionless variable $x_h$, as defined by Eq.~\ref{Pxheq}.}
\end{figure}

The central mass density, $\rho_c\equiv \rho_g(r=0)$, is updated after updating $T_B$, $r_h$, and $\rho_g(r_h)\equiv \rho_B$.  The updated $\rho_c$ value is found by ensuring that the following mass conservation argument is met:
\begin{equation}\label{mgh1}
m_{gh}=f_{mh}m_g = \int_{0}^{r_h} \rho_g(r) \cdot 2\pi r \cdot h \cdot dr,
\end{equation}
where $\rho_g(r)$ is given by Eq.~\ref{rhogr}.  If the expressions given for $T_g(r)$ and $\rho_g(r)$ in Eqs.~\ref{Tgr} and \ref{rhogr} were permitted to extend to radii beyond $r_h$, then we would have $T_g(r)\to 0$ and $\rho_g(r)\to\infty$ when
\begin{equation}
r = r_h \left( 1- \frac{T_B}{T_c}  \right)^{-\nicefrac{1}{\xi}} \equiv r_*. \label{rstar}
\end{equation}
Using $r_*$ to define the dimensionless radius $x\equiv r/r_*$, the hot spot mass density given by Eq.~\ref{rhogr} becomes
\begin{equation}
\rho_g(r) = \rho_c \left( 1 - x^\xi \right)^{-1}.  \label{rhogrC}
\end{equation}
Evaluating the dimensionless radius $x$ at $r=r_h$ gives
\begin{equation}
x_h\equiv r_h/r_*, \label{xh}
\end{equation}
and thus the total fuel mass in the hot spot region, as given by the right hand side of Eq.~\ref{mgh1}, becomes
\begin{align}
m_{gh} &= \int_0^{x_h} \frac{\rho_c}{\left( 1 - x^\xi \right)}\cdot r_*^2 \cdot 2\pi x \cdot h \cdot dx \\
		&= \rho_c \cdot \pi r_*^2 \cdot h \cdot \int_0^{x_h} \frac{2x}{\left( 1 - x^\xi \right)} \cdot dx \\
		&= \rho_c \cdot \pi \left( \frac{r_h}{x_h} \right)^2 \cdot h \cdot \int_0^{x_h} \frac{2x}{\left( 1 - x^\xi \right)} \cdot dx \\
		&= \rho_c \cdot \pi r_h^2h \cdot \frac{1}{x_h^2} \cdot \int_0^{x_h} \frac{2x}{\left( 1 - x^\xi \right)} \cdot dx.
\end{align}
Since the average mass density in the hot spot region is $\bar{\rho}_h = m_{gh}/(\pi r_h^2 h)$, we have
\begin{equation}
\bar{\rho}_h = \rho_c \cdot \frac{1}{x_h^2} \cdot \int_0^{x_h} \frac{2x}{\left( 1 - x^\xi \right)} \cdot dx. \label{rhohbar}
\end{equation}
Now, from Eqs.~\ref{rstar} and \ref{xh}, we have
\begin{equation}
x_h^\xi = \left( 1 - \frac{T_B}{T_c} \right) = \left( 1 - \frac{\rho_c}{\rho_B} \right), \label{xhxi}
\end{equation}
where we have again applied our isobaric assumption; in this case, $\rho_c T_c = \rho_B T_B \Rightarrow T_B/T_c = \rho_c / \rho_B$.  Solving for $\rho_c$ in Eq.~\ref{xhxi} gives
\begin{equation}
\rho_c = \rho_B \left( 1 - x_h^\xi \right).  \label{rhocC}
\end{equation}
Substituting Eq.~\ref{rhocC} into Eq.~\ref{rhohbar} and solving for $\bar{\rho}_h / \rho_B$ gives
\begin{equation} \label{Pxheq}
\frac{\bar{\rho}_h}{\rho_B} = \frac{1 - x_h^\xi}{x_h^2}  \int_0^{x_h} \frac{2x}{1-x^\xi} \cdot dx \equiv \mathcal{P}(x_h),
\end{equation}
where $\mathcal{P}(x_h)$ is a dimensionless, monotonically decreasing function of the dimensionless variable $x_h$; it is defined over our physically relevant domain of $0 \leq x_h \leq 1$ with a range of $0\leq \mathcal{P}(x_h)\leq 1$ (see Fig.~\ref{Pxh}). Because $\mathcal{P}(x_h)$ is a dimensionless function of a dimensionless argument, we only need to evaluate it once and store the results for continuous use throughout the simulation.  With this done, we take the inverse of $\mathcal{P}(x_h)$, with input argument $\bar{\rho}_h/\rho_B$, to get
\begin{equation}
x_h = \mathcal{P}^{-1}\left( \frac{\bar{\rho}_h}{\rho_B} \right).
\end{equation}
Finally, from Eq.~\ref{rhocC}, we get the central density, $\rho_c$, repeated here for completeness as
\begin{equation}
\rho_c = \rho_B \left( 1 - x_h^\xi \right).
\end{equation}


\begin{thebibliography}{97}
\bibitem [{f0()}]{f0}%
  \BibitemOpen
  \href@noop {} {}\bibinfo {howpublished} {Infinite conductivity can be assumed
  by setting $\beta$ to a very large value, which can be accomplished by
  setting $\tau_r$ to a very small value.}\BibitemShut {Stop}%
\bibitem [{uni()}]{uniformJz}%
  \BibitemOpen
  \href@noop {} {}\bibinfo {howpublished} {Note that if the liner current were
  constant in time and uniformly distributed throughout a static liner wall,
  then this would indicate a steady-state condition where the rate of azimuthal
  flux delivery (by the pulsed-power generator) is equal to the rate of
  azimuthal flux dissipation (by ohmic dissipation), and thus we would have
  $\dot{\Phi}_{\theta l}=0$. Our model does not account for this steady-state
  condition (i.e., in Eqs.~\ref{Pohmexact} and \ref{Pohmapprox}, $P_{\Omega}$
  would be zero for this case, even though ohmic dissipation is occurring).
  However, this steady-state ohmic dissipation is negligible relative to the
  dynamic dissipation discussed in the text because $\delta_{skin}< r_l-r_g$.
  For more on the partitioning of supplied electromagnetic energy into ohmic
  and magnetic channels when the associated electrical currents are
  time-dependent and spatially distributed, see
  Ref.~\onlinecite{Knoepfel}.}\BibitemShut {Stop}%
\bibitem [{tra()}]{transnote}%
  \BibitemOpen
  \href@noop {} {}\bibinfo {howpublished} {Note that within the liner region,
  we do not attempt to radially resolve thermal conduction, radiation
  transport, or ohmic dissipation. Doing so would significantly increase the
  complexity of our model, while not significantly improving the accuracy,
  since our model already captures the overall dynamics of a liner implosion
  (i.e., compression, bulk heating, and acceleration) very well relative to
  full 1D radiation magnetohydrodynamics simulations.}\BibitemShut {Stop}%
\bibitem [{uno()}]{unote}%
  \BibitemOpen
  \href@noop {} {}\bibinfo {howpublished} {Note that if the entire fuel region
  is preheated uniformly, i.e., from $r=0$ to $r_g(t_{ph})$, then only the hot
  spot region will exist.}\BibitemShut {Stop}%
\bibitem [{bre()}]{bremsnote}%
  \BibitemOpen
  \href@noop {} {}\bibinfo {howpublished} {Note that our radiation model is a
  bremsstrahlung model that accounts for dopants and/or contaminants mixed into
  the fuel. However, this radiation model does not account for line radiation,
  which can be substantial for dopants and/or contaminants with mid to high
  atomic numbers.}\BibitemShut {Stop}%
\bibitem [{end()}]{endlossnote}%
  \BibitemOpen
  \href@noop {} {}\bibinfo {howpublished} {Note that this axial energy loss is
  due solely to mass flow out of the imploding region. Our model does not
  account for energy losses due to other axially-directed transport processes,
  e.g., axial thermal conduction.}\BibitemShut {Stop}%
\bibitem [{Bzs()}]{Bzshift}%
  \BibitemOpen
\bibfield  {journal} {  }\href@noop {} {}\bibinfo {howpublished} {We have
  shifted the plots of the normalized magnetic field strengths by $+$0.2 so
  that the lines for the case without the Nernst effect can be seen. Since our
  model assumes $B_z(r)/\rho_g(r)=const.$, the normalized $B_z(r)$ would fall
  directly on top of the normalized $\rho_g(r)$ without this
  shift.}\BibitemShut {Stop}%
\end{thebibliography}

%

\end{document}